\documentclass[AMS,Utopia2COL]{WileyNJDv5} 
\articletype{Original Research} 
\usepackage{amsmath,amstext,amssymb,amsfonts,graphics,graphicx}
\usepackage{listings}%
\usepackage{multicol,multirow}
\usepackage{boites,enumerate}
\usepackage{tabularray}
\usepackage{cleveref}
\crefname{figure}{Figure}{Figures}
\crefname{table}{Table}{Tables}
\crefname{section}{Section}{Sections}
\usepackage{amsfonts}%
\usepackage{mathrsfs}%
\usepackage{textcomp}%
\usepackage{manyfoot}%
\usepackage{algorithm}%
\usepackage{algorithmicx}%
\usepackage{algpseudocode}%
\usepackage{changepage}%
\newcommand{\hlc}[2][yellow]{{%
		\colorlet{foo}{#1}%
		\sethlcolor{foo}\hl{#2}}%
}
\usepackage{pgfplots,pgf-pie}%
\pgfplotsset{compat=1.9}%
\usepackage{xurl}%
\usepackage{perpage}%
\MakePerPage{footnote}
\definecolor{coral}{HTML}{F45B69}
\definecolor{coralD}{HTML}{C84A54}
\definecolor{coralL}{HTML}{F67783}
\definecolor{coralLL}{HTML}{F8939D}
\definecolor{coralLLL}{HTML}{FCCBD1}
\definecolor{tangerine}{HTML}{F58A07}
\definecolor{tangerineD}{HTML}{C97205}
\definecolor{tangerineL}{HTML}{F79F34}
\definecolor{tangerineLL}{HTML}{F9B461}
\definecolor{tangerineLLL}{HTML}{FDDEBB}
\definecolor{aero}{HTML}{5DB7DE}
\definecolor{aeroD}{HTML}{4C94B5}
\definecolor{aeroL}{HTML}{79C4E4}
\definecolor{aeroLL}{HTML}{95D1EA}
\definecolor{aeroLLL}{HTML}{CDEBF6}
\definecolor{oxford}{HTML}{0B2545}
\definecolor{oxfordD}{HTML}{08203A}
\definecolor{oxfordL}{HTML}{374E69}
\definecolor{oxfordLL}{HTML}{63778D}
\definecolor{oxfordLLL}{HTML}{BBC9D5}
\definecolor{teal}{HTML}{1F7A8C}
\definecolor{tealD}{HTML}{196675}
\definecolor{tealL}{HTML}{4893A2}
\definecolor{tealLL}{HTML}{71ACB8}
\definecolor{tealLLL}{HTML}{C3DEE4}
\definecolor{plum}{HTML}{aa3fd1}
\definecolor{l2}{HTML}{B6C7F0}
\definecolor{l1}{HTML}{85A0DD}
\definecolor{l3}{HTML}{B1F3C7}
\definecolor{l4}{HTML}{7FE3A1}
\definecolor{l5}{HTML}{FFDA8F}
\definecolor{l6}{HTML}{FFA08F}
\definecolor{l7}{HTML}{FFCE6B}
\usepackage{tcolorbox}
\tcbset{
	tangerine/.style={
		text width=6cm,
		colback=tangerineLLL,
		colframe=tangerine,
		boxrule=1pt,
		fonttitle=\sffamily,
		left=6pt,
		right=3pt,
		top=3pt,
		bottom=3pt,
		before title={\vspace*{3pt}},
	},
	coral/.style={
		text width=6cm,
		colback=coralLLL,
		colframe=coral,
		boxrule=1pt,
		fonttitle=\sffamily,
		left=6pt,
		right=3pt,
		top=3pt,
		bottom=3pt,
		before title={\vspace*{3pt}},
	},
	aero/.style={
		text width=6cm,
		colback=aeroLLL,
		colframe=aeroD,
		boxrule=1pt,
		fonttitle=\sffamily,
		left=6pt,
		right=3pt,
		top=3pt,
		bottom=3pt,
		before title={\vspace*{3pt}},
	},
	oxford/.style={
		text width=13cm,
		colback=white,
		colframe=oxford,
		coltext=black,
		boxrule=1pt,
		fonttitle=\sffamily,
		left=6pt,
		right=3pt,
		top=3pt,
		bottom=3pt,
		before title={\vspace*{3pt}},
	},
}
\hypersetup{
		colorlinks=true,
		linkcolor=blue,
		citecolor=blue,
		urlcolor=blue,
		breaklinks=true,
}
\AtBeginDocument{%
  \renewcommand{\doi}[1]{\href{https://doi.org/#1}{\textcolor{blue}{doi:\,#1}}}%
}

\makeatletter
\long\def\@makecaption#1#2{%
  \vskip3pt
  {\fontsize{8}{10}\selectfont
    \sbox\@tempboxa{{\bf #1}\quad #2}%
    \ifdim \wd\@tempboxa >\hsize
      {\bf #1}\quad #2\par
    \else
      \global \@minipagefalse
      \hb@xt@\hsize{\box\@tempboxa}%
    \fi}%
  \vskip4pt}
\makeatother

\newcommand{\rev}[1]{#1}

\begin{document}

\title{Central Bank Digital Currencies: Where is the Privacy, Technology, and Anonymity?}

\author[1]{Jeff Nijsse}
\author[2]{Andrea Pinto}

\authormark{}
\titlemark{}

\address[1]{\orgdiv{Department of Software Engineering}, \orgname{RMIT University},
	\orgaddress{\state{Hanoi}, \country{Vietnam}}}

\address[2]{\orgdiv{Systems and Computer Engineering Department}, \orgname{Universidad de Los Andes},
	\orgaddress{\state{Bogotá}, \country{Colombia}}}

\corres{Jeff Nijsse <jeff.nijsse@rmit.edu.vn>}

\presentaddress{RMIT University Vietnam, 521 Kim Ma, Ba Dinh District, Hanoi, Vietnam}

\abstract[Abstract]{In an age of financial system digitisation and the increasing adoption of digital currencies, Central Bank Digital Currencies (CBDCs) have emerged as a focal point for technological innovation. Privacy compliance has become a key factor in the successful design of CBDCs, extending beyond technical requirements to influence legal requirements, user trust, and security considerations. Implementing Privacy-Enhancing Technologies (PETs) in CBDCs requires an interdisciplinary approach, however, the lack of a common understanding of privacy and the essential technological characteristics restricts progress. This work investigates: (1) How privacy can be defined within the framework of CBDCs and what implications does this definition have for CBDCs design? and (2) Which PETs can be employed to enhance privacy in CBDC design? We propose a comprehensive definition for privacy that is mapped to the cryptographic landscape for feature implementation. The research is validated against case studies from 20 current CBDCs. The study shows that comprehensive privacy can be designed in the proposal stage, but that privacy does not reach the launched version of the CBDC. \rev{A failure analysis of abandoned privacy pilots identifies four root causes of this research-to-launch gap: regulatory visibility requirements, computational overhead, liability allocation, and institutional incentives.}}

\keywords{CBDC, Central Bank Digital Currency, Privacy, DLT, Blockchain, Cryptography, Privacy Enhancing Technology}

\maketitle
%
\section{Introduction}\label{sec:intro}
In an era characterised by the digitisation of financial systems and the proliferation of digital currencies, Central Bank Digital Currencies (CBDCs) have emerged as a groundbreaking point for technological innovation. According to data from the Human Rights Foundation, 10 Central Banks (CBs) have CBDCs available to their citizens\footnote{
	Bahamas, 
	China, 
	India, 
	Iran, 
	Jamaica, 
	Kazakhstan, 
	Nigeria, 
	Solomon Islands, 
	Russia, and 
	Thailand.
}~\cite{HRF2025}, and more than 93\% of CBs are in active phases of research \cite{Kosse2023}. \rev{The most recent Bank for International Settlements (BIS) survey confirms this trajectory, with 91\% of responding central banks exploring a retail CBDC, a wholesale CBDC, or both \cite{BIS2025}, and as of May 2026 the Atlantic Council tracks 122 countries actively exploring a CBDC, with 41 pilot projects underway \cite{AtlanticCouncil2026}.} As nations contemplate the adoption of digital currencies, questions surrounding the safeguarding of data and privacy of citizens become a central element in the design of CBDCs.

An illustration of public opinion is provided by a 2023 survey conducted by the Bank of Canada~\cite{BoC2023}, which examined perspectives and preferences regarding a digital Canadian dollar. The survey revealed that when asked about trust in the government's ability to issue a secure digital dollar, 79\% of the participants expressed strong disagreement. Furthermore, in ranking the features by importance, more than half of the respondents identified personal control over their data as their primary concern, with one-third emphasising the significance of the ability for anonymous transactions. A similar consultation by the European Central Bank (ECB) in 2022 found that privacy in payments is the most important feature in a potential digital euro \cite{z45}, and the largest group of respondents in a 2024 survey (41\%) chose privacy protection as the most important characteristic \cite{IMF2024}. A statistical approach comes from a Page-rank analysis of keywords associated with CBDCs, finding that privacy is the most searched term with CBDCs \cite{z46}. Considering the established importance of privacy to citizens, this is a topic that has ``scarcely been researched'' \cite{z34}, rather the focus of R\&D is on payment systems \cite{z29}. One systematic literature review found seven themes associated with CBDCs, however, privacy was not one of them, nor was there any mention of privacy technology \cite{z41}. 

The incorporation of Privacy-Enhancing Technologies (PETs) into CBDCs, however, requires an interdisciplinary approach; 
going beyond technical prerequisites to encompass legal mandates, monetary policies, user trust, and security considerations.

Differing views on what constitutes privacy and its fundamental attributes have impeded progress. For example, from the technical perspective, privacy has primarily addressed technical requirements in isolation, treating them as aftermarket solutions that can be added to the system at any time. Conversely, the legal perspective neglects customer requirements for anonymity and cash-like qualities, erring on the side of transaction surveillance.

\rev{The absence of a common privacy definition is compounded by a structural opacity in the CBDC ecosystem itself. Central banks face an inherent tension between their institutional need for monetary control, regulatory compliance, and oversight on one side, and citizens' demand for cash-like privacy on the other. Disclosure practices reflect this tension: design documents are detailed at the research and proposal stages, yet implementation specifics, such as the concrete cryptographic protocols, are routinely unspecified by the time systems launch. Whether this opacity is deliberate or the by-product of procurement and security practices cannot be established from public documents alone; what can be established is that it is systemic, and that it shifts the burden of verification and advocacy onto citizens. This motivates a design principle argued in this paper: CBDC systems should be auditable and transparent by default, such that privacy claims are verifiable properties rather than assurances.}

This work aims to contribute to the discourse by offering a thorough analysis of privacy as applied to CBDCs through two Research Questions: \textbf{(1)} How can privacy be defined within the framework of CBDCs, and what implications does this definition have for CBDC design? and \textbf{(2)} What techniques, methods, and technologies can be employed to enhance privacy in CBDC design.

The primary contributions of the work are \textbf{(1)} A definition of privacy in the context of CBDCs design that is inclusive of three main stakeholder perspectives: legal and regulatory, technological, and transactional, \textbf{(2)} An overview of the contemporary landscape in cryptographic techniques for CBDC design in consideration of the definition of privacy, \rev{\textbf{(3)} A diagnostic evaluation of the definition and technology framework against 20 CBDC case studies, showing that privacy designed at the proposal stage attenuates by the time projects launch, and \textbf{(4)} A failure analysis identifying four root causes of this research-to-launch gap.} These contributions are applicable to various stakeholders, from policy makers and regulators to researchers and developers.

The remainder of the paper is organised as follows. The methodology is introduced in \cref{sec:method} with inclusion criteria for the literature review. \cref{sec:defn} derives a definition of privacy; \cref{sec:pet} provides an overview of privacy-enhancing technologies, beginning with general encryption, leading to more advanced cryptography. Adjacent PETs are addressed in \cref{sec:other}. Case study analysis of the application of PET by central banks is in \cref{sec:case-studies}, and \cref{sec:disc} discusses the state of CBDCs\rev{, the limitations of the study, directions for future work,} and concluding remarks.

\section{Methodology}\label{sec:method}
This study adapts the Design Science Research Methodology (DSRM) \cite{Peffers2007} combining a literature review with artefact development. The DSRM is suited to sociotechnical fields that blend social systems like central banking with technological application of privacy technology. The methodology has \rev{six} steps: (1) Problem identification, (2) Define objectives of the solution and corresponding research questions, (3) Design and development of the artefact, (4) Demonstration, \rev{(5) Evaluation, and (6) Communication; the final two steps are presented jointly below}.

\subsection{Problem Identification} Privacy is crucial in CBDCs design, yet there's no consensus on its integration among stakeholders. Privacy is frequently overlooked in new proposals, and references to general ``encryption'' as a privacy solution lack detail. The complexity of cryptographic privacy technologies adds to the challenge, complicating the real-world implementation of CBDCs.

\subsection{Define objectives of the solution} The main objective of the research is to have a structured analysis of the privacy concepts and its application in the design of CDBCs. The review begins with literature from the 2019--2023 and is supplemented by secondary sourcing and industry grey literature. \rev{Because CBDC development is fast-moving, the systematic review window is updated with 2024--2026 primary sources (central bank progress reports, BIS surveys and papers, and legislative documents), which inform the case studies (\cref{sec:case-studies}) and discussion (\cref{sec:disc}).} Two research questions are formulated to achieve the objectives:

\textbf{RQ1}: How can privacy be defined within the framework of CBDCs, and what implications does this definition have for CBDC design?

\textbf{RQ2}: Which techniques, methods, and technologies can be employed to enhance privacy in design of CBDCs?

\subsection{Design and development of the artefact} 
The literature review is conducted to capture the state of privacy as applied to CBDCs. The Scopus, IEEE Xplore, and Science Direct databases provide the indexed material with the following filters to arrive at the 67 paper review set.

\begin{enumerate}
	\item Keyword combinations: \{privacy AND central bank AND (digital currency OR CBDC)\}.
	\item The date range is from 2019 to 2023.
	\item Peer review documents consisting of articles, conference papers, reviews, or short surveys. Grey literature is included at a later stage for the Case Studies since industry reports and bank publications are not indexed in academic research.
	\item Papers written in English and available for research purposes.
	\item Elimination of repeated documents---a total of 5 documents are duplicates.
	\item Abstract analysis for primary filtering of documents.
	\item Full-text analysis to select the documents that positively contribute to the review. Here, one document is eliminated for not meeting quality standards.
\end{enumerate}

\rev{Screening yielded 73 papers that proceeded to full-text coding. Five pairs of records were identified as duplicates and consolidated, and one record was excluded at the full-text stage on the authors' assessment that it was not of scholarly quality, giving the final 67-paper review set.

Each paper is coded against a fixed six-field template: (1) the definition of privacy offered, if any; (2) how privacy relates to the paper's principal concepts (anonymity, traceability, auditability, security, and similar); (3) the stakeholders to whom privacy is applied; (4) technologies with the potential to enhance privacy; (5) non-technological strategies; and (6) additional key concepts. Each record was assigned a primary coder (37 and 36 papers respectively) with the coding recorded in a shared spreadsheet; the authors then exchanged sets, so that every record was cross-reviewed by the second author, with the perspective classification (technological, legal/regulatory, or transactional) assigned according to the dominant frame of each paper's privacy discussion and disagreements resolved through discussion. No formal inter-coder reliability statistic was computed, a limitation noted in \cref{sec:limitations}. Grey literature is admitted under stricter criteria than the academic set. Official Central Bank and intergovernmental publications (e.g. BIS, IMF, ECB), legislative records, and established trackers (Human Rights Foundation, Atlantic Council) are preferred; commercial press is used only to corroborate or date events, not as a primary basis for classification.}

The literature review and grey literature analysis inform the development of two key artefacts. Privacy Definition: This is derived by synthesising concepts from the literature review and refining them based on insights from grey literature. Technological Guidelines: These are developed by analysing existing privacy-enhancing technologies in CBDCs\rev{, comparing them across deployment criteria,} and evaluating their effectiveness through case studies.

\subsection{Demonstration} The artefacts are explained in the paper: (1) The development of a new privacy definition (\cref{fig:definition}), and (2) The provision of technological guidelines for enhancing privacy in CBDC design (\cref{fig:layers}\rev{ ~and \cref{tab:pet-compare}}).

\subsection{Evaluation and Communication} The final step involves evaluating the developed artefacts--the privacy definition and technological guidelines--by applying them to case studies in CBDCs. These case studies are selected to represent diverse contexts, such as countries with varying regulatory frameworks, technological capabilities, and privacy priorities. The evaluation focuses on assessing the applicability, effectiveness, and practicality of the artefacts in addressing privacy challenges within CBDC design. Data is collected through analysis of policy documents, technical reports, industry papers, and observations of CBDC implementation processes. \rev{The evaluation additionally examines why privacy proposed at the design stage fails to persist into launched systems, developed as a failure analysis in \cref{sec:failure}.} Lastly, communication of the artefacts is via the medium of text and diagrams.

\section{Privacy}\label{sec:defn}
Privacy in the design of CBDCs plays a pivotal role in upholding the principles of autonomy and individual rights in the digital financial landscape. It acts as a safeguard against unauthorised access and surveillance, as well as the potential misuse of sensitive financial information. Preserving privacy within CBDCs is a cornerstone for implementing digital financial systems. Striking a balance between the benefits of CBDCs and implementing robust privacy protection mechanisms is imperative, and ideally addressed at the design stage, rather than waiting for proofs of concept to identify deficiencies.

\subsection{An Absence of Privacy Definitions}
While the importance of privacy in the design of CBDCs cannot be overstated, the first trend in the literature is a gap in the comprehensive understanding of privacy. Studies struggle with the concept of privacy in CBDCs, and a common issue arises---there is not a clear and well-accepted definition. Nearly half of the papers, 33 of 67, do not define privacy in the context of CBDCs or articulate its contextual use. 

Even when definitions are presented, they often fall short of encompassing the multidimensional nature of privacy within CBDCs, only introducing privacy as relates to the domain in question. For example, one definition says privacy is upheld by enabling ``selective disclosure of information'' (the systemic capability) according to the General Data Protection Regulation (GDPR)~\cite{z20}; another that privacy is an individual's perceived risk of loss when disclosing information~\cite{z29}; and another that privacy means upholding the privacy of commercial bank transactions when being audited by central banks~\cite{z30-53}.

\begin{figure}[ht]
	\centering
	\begin{tikzpicture}
		\sffamily
		\pie[
		pos={3,0},
		radius=3,
		color={oxfordLL, aeroLL, coralL, tangerineL},
		rotate=180,
		explode = {0.15, 0, 0, 0}
		]{  
			49.2/\textcolor{white}{{49.2}} {No definition of privacy provided (33)},
			6.0/{\parbox{2cm}{Transact- \\ional (4)}},
			14.9/{Legal (10)},
			29.9/{\parbox{2cm}{Techno-\\logical (20)}}
			}
	\end{tikzpicture}
	\caption{Four main trends emerge in the literature defining privacy. The strongest trend is an absence of a privacy definition found in 33 of 67 papers, followed by a focus on three perspectives: the technological perspective, legal perspective, and transactional perspective.\label{fig:lit-analysis2}}
\end{figure}
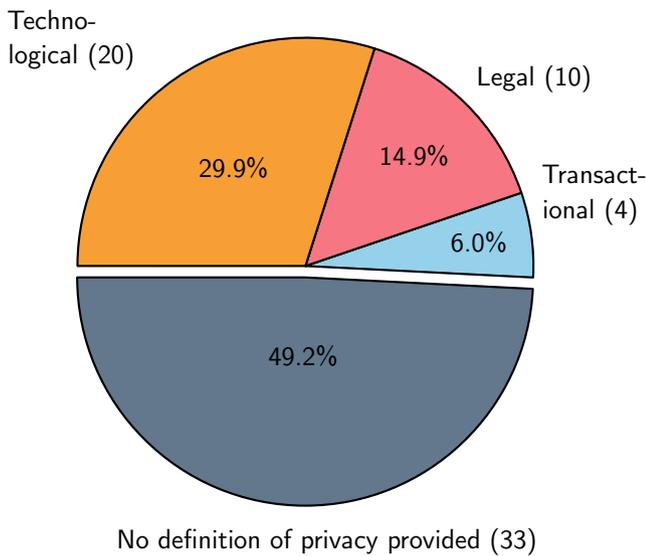

Three main perspectives emerge when viewing privacy: the technological perspective, the legal perspective, and the transactional perspective, as shown in \cref{fig:lit-analysis2}.

The second prevailing trend identified is that of privacy from a technical perspective. In this context, privacy is considered a requirement that can be fulfilled by using a technology. For instance, Distributed Ledger Technology (DLT) is a common element in nearly every design proposal from academia as a method to ensure privacy; blockchain technology being one of the most popular examples of DLT \cite{z01-64,z03,z06,z09-67,z11,z16,z17,z19,z20,z29,z30-53,z36}. 

Blockchains provide transparency, giving participants in the network an open and censorship-resistant ledger of transactions. Using public networks present a trade-off in the quest for transparency, which raises privacy concerns~\cite{z10-49}. Users and data are visible to all participants, which may not align with data protection regulations or personal privacy expectations \cite{z57,z61}. 

However, depending on the architecture, a DLT can or may not achieve a balance between privacy, transparency, and auditing capacity. Thus, privacy-focused technologies, such as privacy coins and Zero-Knowledge Proofs (ZKPs; \cref{sec:zk}), are beginning to address these concerns by providing mechanisms for selective disclosure. These technologies enable users to share specific information based on their need and required disclosure thresholds.


Nevertheless, the questions of: What information is private? What needs protection? From whom should it be shielded? and Who can have access? are questions that require input from multiple stakeholders, not just engineers as enablers of technology.

The third trend in privacy definitions is focused on addressing legal requirements. This perspective presents an additional challenge, for example, designing a CBDC for implementation in a country of the European Union (EU) \cite{z20,z34} differs from proposing one for development in an Asian nation \cite{z37,z54}. In the EU, the GDPR is mandatory as it ensures the protection of users' personal data and privacy rights. However, the approach of the Chinese government may prioritise robust measures for detecting illegal financial activities over personal data protection \cite{z31}. Moreover, there are variations in regulatory compliance even among countries within the same region due to their own understanding of privacy \cite{z51}.

\rev{These regulatory differences are not merely administrative; they rest on divergent philosophical conceptions of what privacy is and whom it serves. The liberal-democratic tradition treats privacy as an individual right held against the state and other actors---from the early articulation of a ``right to be let alone'' \cite{WarrenBrandeis1890} to Westin's influential framing of privacy as the individual's claim to determine when, how, and to what extent information about them is communicated \cite{Westin1967}. The GDPR is a direct descendant of this tradition. A state-centric conception, by contrast, subordinates individual informational control to collective goals such as financial stability, social order, and crime prevention; the eCNY's design, in which anonymity is \textit{managed} by the issuing authority rather than held by the user, is the canonical CBDC expression of this view. The two conceptions assign opposite default owners to transaction data, and consequently no single accepted privacy framework can be universal. The definition developed in this paper does not attempt to resolve this philosophical divergence; instead it deliberately parameterises it; terms such as \textit{right or request} and \textit{adequate balance} are jurisdiction-dependent by design, while the layered technology stack (\cref{fig:layers}) remains common to both conceptions.}

Despite the differences, there is a necessity to align CBDC designs with the legal requirements of various stakeholders, including governments, financial regulatory institutions, and auditing bodies. Surprisingly, defining privacy is still a challenge even for policymakers. Although documents such as GDPR outline specific principles and rules regarding the handling and protection of personal data, it doesn't provide an explicit definition of privacy. However, to bridge the gap between the regulatory landscape and the technical design of CBDCs, a comprehensive definition of privacy that incorporates both legal and technical privacy requirements can serve as a step toward the development of privacy focused CBDCs.

Finally, the least common trend in defining privacy looks at the transactional perspective \cite{z15,z21,z38,z39}. This perspective is based on protection of data associated with a financial transaction such as: digital identities of senders and recipients, transactional amount and ID, fees, and transactional status. This perspective asks for a high degree of untraceability as part of the characteristics of CBDCs. Current digital transactions offer complete traceability by central and retail banks. Those seeking privacy or anonymity often turn to cash as their preferred option. Untraceability, while associated with a high level of privacy, must be balanced with the need for legal compliance and the prevention of illicit use. This requires a certain degree of auditing capacity to be integrated into CBDC designs. If CBDCs are to be a key part of the financial system, there are ancillary data within the transactional perspective used by supporting industries including credit and tax agencies that need to be included in privacy considerations.

\subsection{Essential Privacy Concepts}\label{sec:concepts}
\begin{figure}[ht]
	\centering
	\includegraphics[trim={10mm 2mm 1mm 2mm}, clip, width=0.5\textwidth]{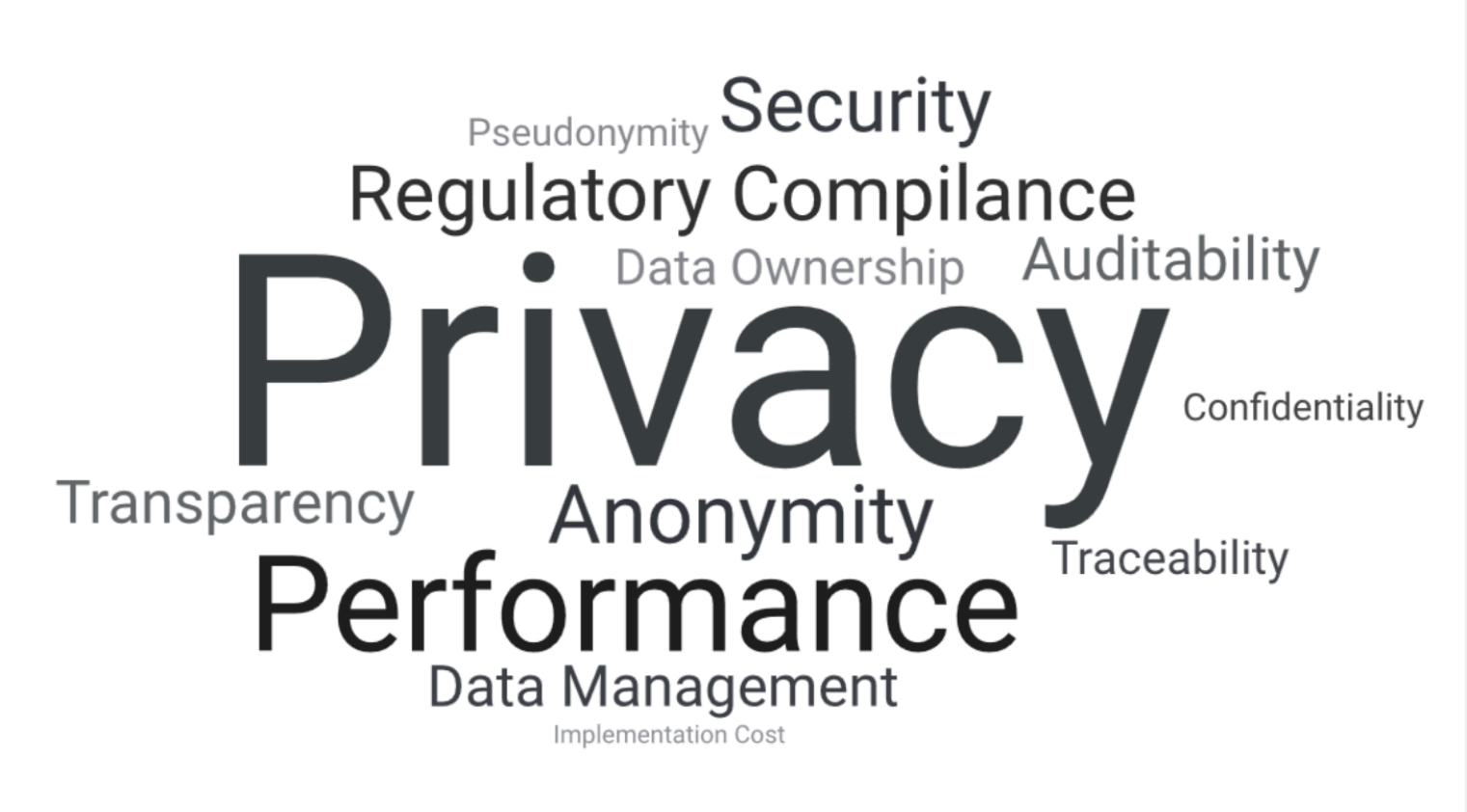}
	\caption{Privacy related concepts in the context of CBDCs design. The concepts most frequently associated with privacy, listed in descending order, include: performance, regulatory compliance, anonymity, security/cybersecurity, and data management.}\label{fig:word-cloud2}
\end{figure}

In CBDC design, privacy acts as the central thread, weaving together security, compliance, performance, transparency, anonymity, traceability, data ownership, and data management into a financial system (\cref{fig:word-cloud2}). Difficulties in pinning down a definition of privacy arise from the interconnected nature of these concepts.

Privacy and security are often interrelated \cite{z04-62,z05,z08,z09-67,z12,z24,z27,z30-53,z33,z45,z50,z52,z54,z60,z69}, but privacy should be distinguished from confidentiality, a well-known property of information security that primarily deals with protecting information from unauthorised access. On the other hand, privacy forms a barricade against inappropriate use by authorised entities, ensuring the sanctity of user information. Second, regulatory compliance relies on the privacy of financial data \cite{z09-67,z10-49,z11,z15,z19,z26,z33,z37,z38,z50,z56}. Privacy safeguards not only protect the individual but also support trust in the regulatory bodies, enabling the seamless operation of CBDCs within the confines of the law \cite{z09-67,z10-49,z20,z26,z44}. Third, transparency and anonymity coexist in the privacy landscape. Transparency assures users that the system is trustworthy, while anonymity empowers them to transact without fear of undue scrutiny \cite{z07,z10-49,z21,z24,z38,z42,z43,z50,z51,z66}. Meanwhile, traceability bridges these two facets by enabling authorised entities to track illicit activities without infringing on the privacy of law-abiding users. Fourth, data ownership and management \cite{z01-64,z11} is tied to privacy where individuals retain control (or not) over their financial data, and the management thereof.

Analysing the connections between these concepts and privacy cannot occur in isolation, as various relevant parties have competing interests that must be considered.
\subsection{Stakeholders in the design of CBDCs}\label{sec:stakeholders}
\begin{figure*}[htb]
	\centering
	\begin{tikzpicture}
		\sffamily
		\begin{axis}[
			width=0.8\textwidth,
			height=8cm,
			xbar,
			y axis line style = { opacity = 0 },
			axis x line       = none,
			tickwidth         = 0pt,
			enlarge y limits  = 0.18,
			enlarge x limits  = 0.25,
			bar width=14pt, 
			symbolic y coords = {
				Malicious Actors,
				Businesses,
				Governments,
				Commercial Banks/ Financial Institutions,
				Regulatory/Auditing Institutions,
				Customers/Citizens/Users
			},
			nodes near coords,
			every node near coord/.append style={
				font=\normalsize
			},
			xlabel style={font=\normalsize},
			ylabel style={font=\normalsize}, 
			yticklabel style={
				align=right, 
				inner sep=-15pt, 
				text width=4.3cm, 
				font=\normalsize
			}
			]
			\addplot[fill=oxford, draw=none] coordinates {
				(6,Malicious Actors) 
				(7,Businesses) 
				(16,Governments) 
				(26,Commercial Banks/ Financial Institutions) 
				(36,Regulatory/Auditing Institutions) 
				(49,Customers/Citizens/Users)};
		\end{axis}
	\end{tikzpicture}
	\caption{Histogram showing the stakeholders in CBDC privacy. The most important stakeholders are customers, followed by financial regulations or auditing institutions. The counts are not mutually exclusive.}
	\label{fig:Stakeholders}
\end{figure*}
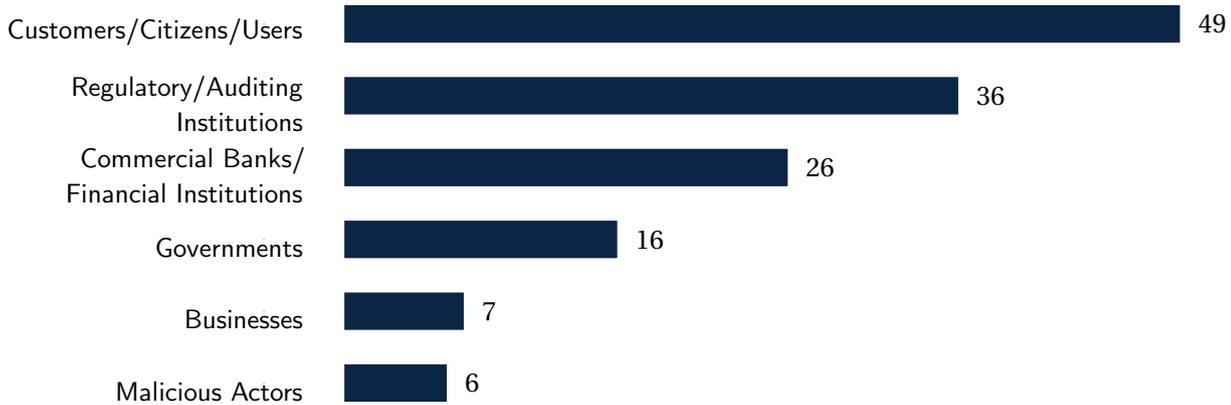

To protect users' privacy in the design of CBDCs is a task that needs a collective effort of diverse stakeholders. \cref{fig:Stakeholders} identifies the stakeholders in CBDC design by frequency weighting: (1) Citizens as the primary user act as customers of the CB, and thus to build trust and catalyse adoption the needs of the customer must be accounted and provided for \cite{z01-64,z04-62,z06,z38}; (2) Financial regulators and auditing institutions, with their regulatory standards, guide the development of privacy-compliant CBDCs \cite{z05,z07,z16}; (3) Commercial banks and financial institutions, in charge of managing and securing sensitive financial data \cite{z46,z59,z63}, (4) Government authorities, entrusted with the formulation and enforcement of laws, hold the keys to the legal framework within which CBDCs must operate, ensuring that they protect user privacy while complying with regulations \cite{z17,z18,z22}, and (5) Businesses, as a specific type of CBDC user \cite{z31,z52,z54}, are not only obliged to adhere to privacy regulations but also play a pivotal role in system implementation to demonstrate trust for their customers which are citizen consumers. Lastly, malicious actors, which are individuals or organisations with the intent of engaging in illegal activities such as, fraud, and money laundering within the CBDC system \cite{z11,z57,z50}.

There are additional stakeholders not explicitly mentioned that warrant consideration to safeguard privacy. Privacy advocates and civil society organisations acting on behalf of citizens are the guardians of civil liberties monitoring the protection of user privacy rights; technical developers and engineers are architects who can convert privacy principles developed by governments and regulatory institutions into operational designs; and academics and researchers to enrich and disseminate the discourse with their ideation and critical evaluation.

\rev{The frequency weighting in \cref{fig:Stakeholders} documents how often the literature attends to each group, but attention is not influence. Mapping the same stakeholders onto a power--interest grid (\cref{fig:power-interest}) exposes an asymmetry that the frequency count conceals: the groups with the greatest decision-making power over CBDC design (central banks, governments, and regulators) are those whose institutional interest favours oversight, while the group with the most acute interest in transactional privacy (citizens) holds the least direct power. Citizens' influence is largely exercised indirectly, through adoption refusal and public opinion---a channel that is nevertheless potent: negative public sentiment was a contributing factor in the Bank of Canada shelving its digital dollar \cite{Canada2024}, and surveillance concerns drove the legislative prohibition of a retail CBDC in the United States \cite{HR1919}. Above the national actors sit the supranational standard-setters---most prominently the Financial Action Task Force (FATF), whose AML/CFT standards bind national design choices regardless of jurisdiction, alongside the BIS and IMF as conveners of CBDC experimentation; \cref{sec:failure} shows their requirements operating as design non-negotiables. Commercial banks hold high power and a strong but ambivalent interest: customer transaction data is a commercial asset, yet two-tier architectures assign them the data-protection liability, so they tend to favour privacy from third parties but not from themselves. Privacy advocates hold high interest but low formal power, acting mainly by shifting public opinion, and malicious actors, while holding no legitimate power, exert anticipatory influence: the supervisory apparatus that constrains every other actor is constructed in expectation of them. Notably, the high-power/low-interest quadrant is empty: no powerful actor in the CBDC ecosystem is indifferent to the privacy outcome, which is why privacy cannot be treated as an aftermarket design add-on. The grid is a design-time snapshot: citizens' power is largely latent and surfaces at adoption, and individual jurisdictions shift position with their philosophical orientation toward privacy (\cref{sec:defn}). This power--interest view explains a pattern developed in \cref{sec:case-studies}: where privacy survives into deployed systems, it is typically the privacy that powerful stakeholders need (data security, tiered liability), not the privacy citizens ask for (transactional anonymity).}

\begin{figure}[htb]
	\begin{tikzpicture}[scale=1.1]
		\sffamily
		\draw[->, thick] (0,0) -- (7.2,0);
		\node[font=\normalsize] at (3.6,-0.4) {Stake in privacy outcome};
		\draw[->, thick] (0,0) -- (0,6.2);
		\node[font=\normalsize, rotate=90, inner sep=1pt] at (-0.28,3.1) {Power over design};
		\draw[dashed, gray] (3.6,0) -- (3.6,6);
		\draw[dashed, gray] (0,3) -- (7,3);
		\node[fill=oxfordLLL, rounded corners=2pt, inner sep=3pt, font=\footnotesize, align=center] at (6.2,5.7) {Standard Setters\\(FATF, BIS)};
		\node[fill=oxfordLLL, rounded corners=2pt, inner sep=3pt, font=\footnotesize, align=center] at (4.9,4.9) {Central Banks/\\Governments};
		\node[fill=oxfordLLL, rounded corners=2pt, inner sep=3pt, font=\footnotesize, align=center] at (6.3,4.1) {Regulatory \&\\Auditing Bodies};
		\node[fill=aeroLL, rounded corners=2pt, inner sep=3pt, font=\footnotesize, align=center] at (4.5,3.5) {Commercial\\Banks};
		\node[fill=coralLL, rounded corners=2pt, inner sep=3pt, font=\footnotesize, align=center] at (6.0,1.4) {Citizens/\\Users};
		\node[fill=coralLL, rounded corners=2pt, inner sep=3pt, font=\footnotesize, align=center] at (4.4,0.85) {Privacy\\Advocates};
		\node[fill=tangerineLL, rounded corners=2pt, inner sep=3pt, font=\footnotesize, align=center] at (2.0,1.6) {Businesses};
		\node[fill=tangerineLL, rounded corners=2pt, inner sep=3pt, font=\footnotesize, align=center] at (3.8,2.3) {Developers/\\Academics};
		\node[draw=black, fill=gray, text=white, rounded corners=2pt, inner sep=3pt, font=\footnotesize, align=center] at (6.3,0.4) {Malicious Actors};
	\end{tikzpicture}
	\caption{\rev{Power--interest grid of CBDC privacy stakeholders. The horizontal axis measures the magnitude of an actor's stake in the privacy outcome, not support for privacy. Malicious actors hold no legitimate power but exert anticipatory influence. Positions are the authors' synthesis of the case-study record.}\label{fig:power-interest}}
\end{figure}

Together, these stakeholders must work to implement privacy preservation aspects within the CBDCs landscape, under the guidance of the central banks that oversee national monetary policies, and together shoulder the responsibility of balancing financial control and individual privacy.

\subsection{A Definition of Privacy in the Context of CBDCs}
\begin{figure*}[htb]
	\centering
	\begin{tikzpicture}[auto,
		arrow/.style = {->, ultra thick, shorten >=-3.3pt, shorten <=-4.1pt},
		]
		\sffamily
		\node[tangerine, anchor=north west] (tx) at (0,0) {
			\begin{tcolorbox}[title={Transactional System Perspective}, width=\textwidth, tangerine]
				 Privacy as the ability of the system to keep the information of parties involved in the transaction untraceable.
			\end{tcolorbox}
		};
		
		\node[coral, anchor=north west] (legal) at (7,0) {
			\begin{tcolorbox}[title={Legal \& Regulatory Perspective}, width=\textwidth, coral]
				 Privacy of personal data as the right and/or ability of an individual to control sensitive data that is possessed by other parties and to keep information private.   
			\end{tcolorbox}
		};
		
		\node[oxford, anchor=north west] (main) at (0,-3.3) {
			\begin{tcolorbox}[title={CBDC Privacy Definition}, width=\textwidth, oxford]
				Privacy in the context of CBDCs is the \hlc[aeroLLL]{protection of personal data and financial information} fostering \hlc[tangerineLLL]{security and confidentiality in a digital financial ecosystem}. Privacy encompasses the \hlc[coralLLL]{right or request of CBDC end-users to control access to their transactional data, maintain anonymity to a desired extent}, and \hlc[tangerineLLL]{safeguard} their digital identities against \hlc[aeroLLL]{unwarranted intrusion or illegal surveillance}, while \hlc[coralLLL]{ensuring compliance with regulatory measures} to strike an adequate balance between financial integrity and personal freedom.
			\end{tcolorbox}
		};
		
		\node[aero, anchor=north west] (tech) at (3.5,-8) {
			\begin{tcolorbox}[title={Technological Perspective}, width=\textwidth, aero]
				 Privacy as the protection against unintended disclosure of identity data and transaction information.    
			\end{tcolorbox}
		};
		\draw[arrow] (tech) -- (main);
		\draw[arrow] (legal) -- (main);
		\draw[arrow] (tx) -- (main);
	\end{tikzpicture}
	\caption{Privacy defined in the context of a CBDC that integrates the three perspectives of the transactional, legal, and technological.\label{fig:definition}}
\end{figure*}
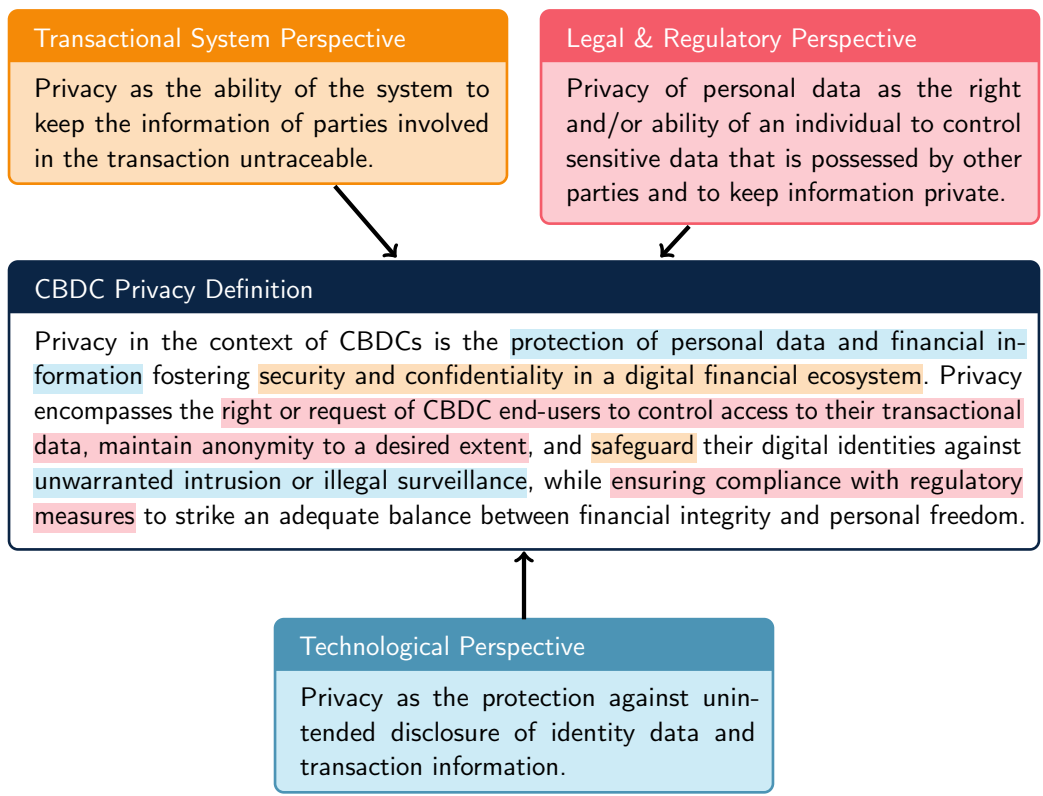

The definition is derived from the privacy concepts, stakeholders, and from the legal, transactional, and technological perspectives, and shown visually in \cref{fig:definition}.

Privacy in the context of CBDCs is: \textit{the protection of personal data and financial information fostering security and confidentiality in a digital financial ecosystem. Privacy encompasses the right or request of CBDC end-users to control access to their transactional data, maintain anonymity to a desired extent, and safeguard their digital identities against unwarranted intrusion or illegal surveillance, while ensuring compliance with regulatory measures to strike an adequate balance between financial integrity and personal freedom.}

The definition serves as a broad foundation, acknowledging the multifaceted nature of the subject. It emphasises the importance of prioritising privacy in the design of CBDCs, recognising that achieving this goal goes beyond creating a theoretically perfect model. It remains flexible with respect to the interpretation of subjective terms \textit{right or request}, and \textit{adequate balance}, while highlighting their importance by fact of inclusion. Further, terms such as \textit{digital ID} and \textit{personal freedom} can be refined, each according to their jurisdiction.

To develop practical and real-world CBDC proposals, the technological perspective must align not only with legal requirements but also with the complexities of the broader financial system, extending far beyond the realm of central banks. Governments can have varying priorities in CBDC design, with some emphasising legal requirements that may limit technological capabilities, while others may be constrained by limited access to high-tech solutions. The question of enabling the appropriate balance between these perspectives should be addressed by designers in the initial phases of CBDCs design.

\rev{The definition becomes operational when each legal obligation is traced to a concrete design choice. Four mappings recur. (1) Data-protection law: the GDPR's data-minimisation principle implies that intermediaries should hold only the data strictly necessary to process a payment, which in design terms favours blind signatures (\cref{sec:blind}) and ZKP-based selective disclosure (\cref{sec:zk}) over plaintext account records. (2) The GDPR right to erasure conflicts directly with append-only ledgers, implying either off-chain storage of personal data with on-chain commitments (\cref{sec:pc}), or a non-DLT architecture. (3) AML/CFT obligations impose transaction-monitoring thresholds, which map onto tiered wallet designs---low-value tiers with reduced KYC and value caps (as in the Sand Dollar and eNaira), or anonymity vouchers for limited amounts (as proposed for the digital euro \cite{ECB2019})---rather than onto blanket surveillance. (4) Auditability requirements map onto accountable-anonymity primitives such as BBS+ group signatures with a designated opener (\cref{sec:bbs}) and auditable commitments, rather than onto unconditional anonymity. Each mapping constrains the PET selection developed in \cref{sec:pet}, and \cref{tab:pet-compare} evaluates the candidate technologies against  these regulatory criteria.}


\section{Privacy Enhancing Technology}\label{sec:pet}
Privacy Enhancing Technologies encompass a diverse array of tools and methodologies designed to safeguard personal information in a digital environment. These technologies range from fundamental protective measures, such as password protection for documents, to advanced cryptographic protocols that enable users to preserve their anonymity online.

The integration of PETs into digital frameworks has become increasingly vital as the demand for privacy protection intensifies in our interconnected world. This necessity is particularly evident in the context of CBDCs, where the interplay between blockchain and cryptographic methods plays a crucial role in addressing privacy concerns. As blockchains are designed for transparency, their open ledger exposes transaction metadata that can compromise user privacy, highlighting the need for cryptographic methods to enhance privacy.

\begin{figure}[htb]
	\centering
	\begin{tikzpicture}
		\sffamily
		\begin{axis}[
			width=0.45\textwidth,
			height=5cm,
			xbar,
			y axis line style = { opacity = 0 },
			axis x line       = none,
			tickwidth         = 0pt,
			enlarge y limits  = 0.2,
			enlarge x limits  = 0.25,
			bar width=14pt, 
			symbolic y coords = {
				Secure Hardware,
				Digital Identity,
				Blockchain/DLT,
				Cryptography
			},
			nodes near coords,
			every node near coord/.append style={
				font=\normalsize
			},
			xlabel style={font=\normalsize},
			ylabel style={font=\normalsize}, 
			yticklabel style={anchor=east, yshift=3pt, align=right, inner sep=-25pt, text width=4cm, font=\normalsize}
			]
			\addplot[fill=oxford, draw=none] coordinates {
				(1,Secure Hardware)
				(3,Digital Identity) 
				(20,Blockchain/DLT) 
				(30,Cryptography) 
			};
		\end{axis}
	\end{tikzpicture}
	\caption{Histogram of Privacy-Enhancing Technology as applied to CBDCs derived from the literature involves applying traditional cryptographic methods but also suggests privacy can be met by applying blockchain technology. Digital identity and secure hardware form and a minority of applications. The counts are not mutually exclusive. \label{fig:categories}}
\end{figure}
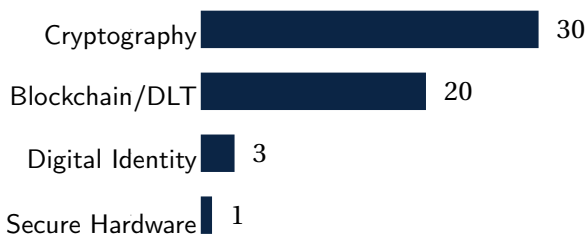

There is a dominance between blockchain and cryptography as the general tools available to designers, seen in \cref{fig:categories}, where 58\% of studies imply that privacy in CBDC development can be met by the introduction of technical methods. The level of understanding and detail comes in varying degrees. Three sources say nothing more than a general sense that cryptography can be used to assist with privacy \cite{z13,z45,z54}, while a further six offer no more insight than mentioning the term `encryption' \cite{z04-62,z12,z31,z59,z65,z72}. This leaves a significant void of details to help practitioners.

\begin{figure*}[htb]
	\centering
	\begin{tikzpicture}
		\sffamily
		\begin{axis}[
			ybar=0pt,
			/pgf/bar shift=-3pt,
			width=\textwidth,
			height=6cm,
			bar width=13pt, 
			tickwidth  = 0pt,
			ylabel={Mentions},
			yticklabels={},
			y axis line style = { opacity = 0 },
			axis x line*=bottom,
			tick label style={font=\footnotesize},
			x tick label style={rotate=45, 
				anchor=east, 
				align=right},
			xtick={1,...,14},
			xticklabels={
				{ZK\phantom{.}},
				{Homomorphic\phantom{.}},
				{Pedersen\phantom{.}},
				{Blind Sig.},
				{ElGamal\phantom{.}},
				{MPC\phantom{.}},
				{Ring Sig.},
				{Digital Sig.},
				{Schnorr Sig.},
				{BBS+Sig.},
				{Groth Sig.},
				{Mixing\phantom{.}},
				{VRF\phantom{.}},
				{Hardware\phantom{.}}},
			nodes near coords,
			nodes near coords style={text=black},
			every node near coord/.append style={
				font=\fontsize{8pt}{10pt}\selectfont
			},
			every axis plot/.append style={fill},
			ymin=0,
			enlarge x limits=0.04,
			]
			\addplot[fill=oxford, draw=none] coordinates {
				(1,14)
				(2,5)
				(3,5)};
			\addplot[fill=tangerine, draw=none] coordinates {
				(4,4)};
			\addplot[fill=oxford, draw=none] coordinates {
				(5,3)
				(6,3)};
			\addplot[fill=tangerine, draw=none] coordinates {
				(7, 3)    
				(8, 1)
				(9, 1)
				(10,1)
				(11,1)};    
			\addplot[fill=oxford, draw=none] coordinates {
				(12,1)
				(13,1)
				(14,1)
			};
		\end{axis}
	\end{tikzpicture}
	\caption{Cryptographic techniques sorted by occurrence in the privacy literature in the context of CBDCs. Zero-knowledge proofs are the most common with reference in 14 of 67 sources. Digital signature variants are in \hlc[tangerineLLL]{orange} and form, in aggregate, the next most prominent method with 11 references. The mentions are not mutually exclusive.\label{fig:techniques}}
\end{figure*}
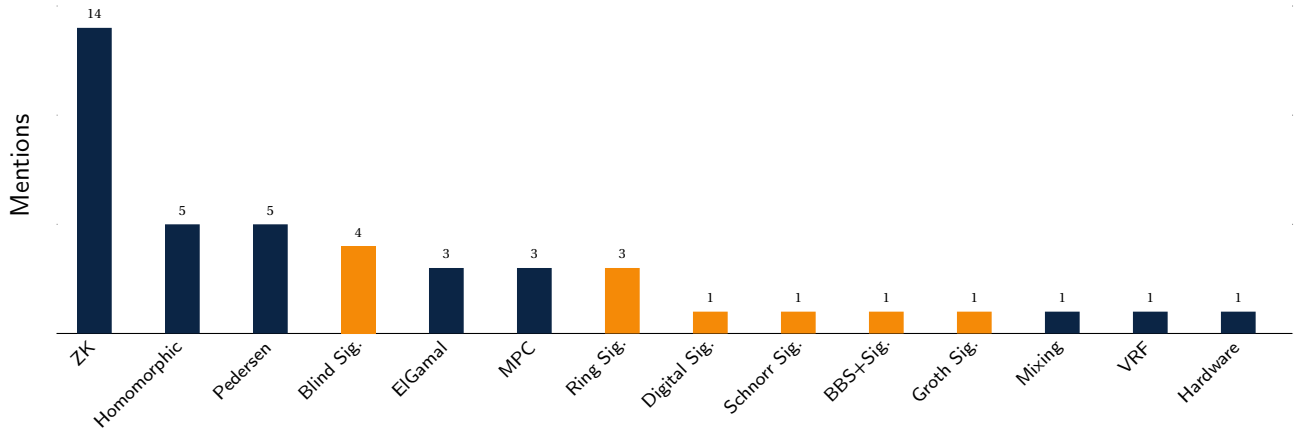

The specific cryptographic methods are broken down by count in the literature in \cref{fig:techniques}. Zero Knowledge Proofs dominate the discussion followed by digital signatures. The long tail shows singular mentions of specific signature techniques along with coin mixing, verifiable random functions, and secure hardware.

From the definition in \cref{fig:definition}, privacy can be organised into layers that build on each other. Privacy, in the digital environment requires: (A) Protection of personal data, (B) Protection of financial information, (C) The right of users to maintain anonymity, (D) The right or request of CBDC users to control access to their transaction data, and (E) Ensuring compliance with regulatory measures.

All cryptographic methods described in \cref{fig:techniques} are discussed presently and integrated with the definition of privacy in a layered stack shown in \cref{fig:layers}.
\subsection{Encryption}
The layered approach begins with the protection of data seen as Layer A in \cref{fig:layers}. Encryption safeguards data by converting it into an unreadable format, which can only be restored to its original state using a specific key. Consequently, only those who possess this key have the ability to access the decrypted information. When Alice deposits money in her bank, she has the expectation that only the bank (and herself) can access the account. In a CBDC context, these data can be transaction data, account data, access keys, Know Your Customer (KYC) data, and any associated metadata. Regardless of the type of data, standard encryption methods apply. Public-key cryptosystems provide the common method used to secure data, for example, when sending credit card information online\footnote{The most well-known is RSA encryption \cite{RSA1978} that allows for encryption and digital signatures by an asymmetric key pair. In practice RSA is used to establish a session key which is then used for data transfer via AES or similar.}, or in encrypted email services. 


Kiayias et al. \cite{z51} propose a private and regulated CBDC with threshold ElGamal encryption. (Their scheme also involves Blind signatures (\cref{sec:blind}) and ZKPs (\cref{sec:zk}).) The ElGamal cryptosystem~\cite{ElGamal1985} is a public-key cryptosystem that builds upon the principles of the Diffie-Hellman key exchange. Apart from its use as an encryption scheme, ElGamal has been the foundation for various digital signature schemes. ElGamal has a probabilistic element from random number inputs that makes it semantically secure, meaning that even if the same plaintext is encrypted multiple times with the same public key, the resulting ciphertexts will appear random and unrelated, making it difficult for an attacker to gain insights into the plaintext based on ciphertext patterns alone. Enhancing another potential feature of CBDCs, Chu et al. \cite{z74} suggest ElGamal can ensure user privacy in offline e-cash, however, it is unclear how this uniquely benefits offline peer-to-peer transactions. Another CBDC design~\cite{z05} suggests that the central bank encrypts user KYC data with the ElGamal cryptosystem, however, this is merely standard encryption already in use by banks and data custodians.

Standard encryption can digitally protect user data, Layer A, but to protect transaction data additional functionality is required. A transaction is an exchange of information between two parties, which can be stated as: \textit{Alice sends Bob 1 coin}. In this transaction, it does not help for Alice to encrypt her coin, as Bob has no assurance that the encrypted object is now his coin. Additionally, were Alice to share her key so Bob can decrypt the object, then Alice has compromised her own security.\footnote{Note the difference between the bank decrypting Alice's information where Alice trusts the bank.} Both parties need to have access to, or to ``see'', the transaction to agree on the details. Digital signatures solve this problem. 

\begin{figure*}[ht]

		{\normalsize\sffamily
			\begin{tblr}{
					colspec = {@{}
						X[2.65,l,m,rightsep=3pt]
						X[c]
						X[c]
						X[c]
						X[c]
						X[c]
						X[c]
						X[c]
						X[c]
						@{}
					},
					cell{6}{2} = {l1}, 
					cell{5}{2} = {l2},      
					cell{4}{2,3,4} = {l2},  
					cell{2}{2} = {l2},      
					cell{2}{4} = {l4},      
					cell{2}{6} = {l5},      
					cell{2}{8} = {coralLL},      
					cell{3}{6} = {l3}, 
					cell{4}{6} = {l3}, 
					cell{3}{2} = {l4}, 
					cell{4}{5} = {l4}, 
					cell{4}{7} = {l5}, 
					cell{4}{9} = {coralLL},
					cell{4}{8} = {l7},
					hline{2,3,4,5,6,7},
					rowsep=0pt,
					colsep=0pt,
					rows = {14.5mm},
					rowspec = {Q[6mm]QQQQQ}
				}
				\textbf{{\normalsize Layer}} & \SetCell[c=8]{c,m}{\textbf{{\normalsize Cryptographic Tools}}} \\
				\textbf{(E)} Ensuring compliance with \phantom{\textbf{(X)~}}regulatory measures
				& \SetCell[c=2]{m}{BBS+\\Signatures}& 
				& \SetCell[c=2]{m} HE &
				& \SetCell[c=2]{m} {Pedersen\\Commitments (PC)} &  
				& \SetCell[c=2]{m} {Privacy\\Pools} &\\
				\textbf{(D)} Right or request to \phantom{\textbf{(X)~}}control access to \phantom{\textbf{(X)~}}transaction data &  \SetCell[c=4]{c} Homomorphic Encryption (HE) &&&& \SetCell[c=4]{c} Zero-Knowledge Proofs (ZKP)&&&\\
				\textbf{(C)} Right of users to \phantom{\textbf{(X)~}}maintain anonymity
				&\parbox[c][14.5mm]{14.5mm}{\begin{center}Blind\\Signatures\end{center}}
				&\parbox[c][14.5mm]{14.5mm}{\begin{center}Ring\\Signatures\end{center}}
				&\parbox[c][14.5mm]{14.5mm}{\begin{center}Schnorr\\Signatures\end{center}}
				& HE 
				& ZKP 
				& PC 
				& MPC  
				&Mixing \\
				\textbf{(B)} Protection of transaction \phantom{\textbf{(X)~}}data (Alice + Bob) & \SetCell[c=8]{c} Standard Digital Signatures: RSA, ECDSA\\
				\textbf{(A)} Protection of personal \phantom{\textbf{(X)~}}data (Alice only) & \SetCell[c=8]{c} Standard Encryption: RSA, EC, ElGamal\\
			\end{tblr}
		}

	\caption{The PET cryptographic stack and the available tools, derived from the definition of privacy in \cref{fig:definition}. Begin at Layer A with standard encryption and add layers to meet the design requirements of the CBDC. This is cross referenced with the \rev{Operational Requirements in \cref{tab:pet-compare}} and the Case Studies in \cref{tab:case-studies}.\label{fig:layers}}
\end{figure*}   

\subsection{Digital Signatures}
First proposed by Rivest, Shamir, and Adleman (RSA;~\cite{RSA1978}), digital signatures prove the authenticity and integrity of a message. A signer uses their private key to sign a message. Anyone with the corresponding public key can verify that the message was signed by the holder of the private key and that the message has not been tampered with. The message data in this case is transaction data, where every transaction requires a signature. For Alice to give Bob one coin, she signs the transaction with her private key, which Bob can then verify using Alice's public key. (Alice has not had to reveal her private key.) The Elliptic Curve Digital Signature Algorithm (ECDSA) is a commonly applied scheme used in Bitcoin and Ethereum, while RSA signatures are common in web applications.

In the absence of digital signatures, online trust is fundamentally compromised. This is no different for a CBDC. Systems eschewing digital signatures typically resort to analogue counterparts, like written signature authentication, or verbal tests. In this manner, digital signatures can be a proxy for identity, for example, where a mobile phone can handle transaction signing. 


In Layer B, digital signatures are required for users to transact with others in the network. Myriad variations in digital signatures provide properties that can add privacy to the stack to allow users the right to maintain anonymity.

\subsubsection{Ring Signatures}
Ring signatures are useful in the design of CBDCs~\cite{z10-49,z19,z50}. The key feature of ring signatures is that they allow a signer to create a signature on behalf of a ring of potential signers without revealing the individual signer~\cite{Rivest2001}. This allows individual transactions to hide, or be anonymous within a group. For example, all transactions under a certain threshold can be signed by the group. This ring is a set of public keys, and any holder of the corresponding private key to one of those public keys can create a ring signature. The signer does not need to collaborate or obtain permission from the other members of the ring (or the signer's own keys). This provides ambiguity for the signer. It's like a group signature \cite{Chaum1991} but without any centralised group setup\footnote{In a standard group signature scheme there is a group manager that handles the group public key and oversees revocation in case of malicious activity.}. This allows the sender of a transaction to hide in a crowd and is used in Monero transactions. Importantly, these transactions are not completely anonymous because the ring itself is identifiable; however, they provide a layer of transaction privacy above regular signatures by unlinking the sender. 


\subsubsection{Schnorr Signatures}
Schnorr signatures \cite{Schnorr1991} are a simple and efficient elliptic curve signature scheme. Unlike ECDSA, which requires a unique random number for each signature (risking key exposure if repeated), Schnorr signatures inherently avoid this risk and provide a linear structure allowing for multiple signatures. Their linearity makes them particularly attractive for complex cryptographic protocols and reduces the size of multi-sig transactions. Batch verification also improves efficiency. Privacy can be improved by the key combination properties, for example, participants can combine public keys into a single key which appears as such to the verifier. Bitcoin improvement proposal 340 is to introduce Schnorr signatures for Bitcoin\footnote{\url{https://github.com/bitcoin/bips/blob/master/bip-0340.mediawiki}}. Lie et al. \cite{z05} have an anonymous and traceable CBDC operating in three tiers similar to traditional banking: CBs, commercial banks, and users. Coin creation by the CB is secured with Schnorr and BBS+ signatures.

\subsubsection{BBS+ Signatures}\label{sec:bbs}
Layer E in \cref{fig:layers} is ensuring compliance with regulatory measures and designs can be enhanced here by BBS+ methods. BBS+\footnote{Named after the authors: Boneh, Boyen, and Shacham.} signatures \cite{Boneh2004} are a type of group signature scheme with the feature of being able to reveal the identity of the signer if necessary (through a designated entity or group manager). This allows for accountability while preserving privacy. BBS+ might be more suitable than ring signatures for systems where conditional anonymity is a requirement, such as a banking authority that is subject to legal requirements. Broadly, BBS+ signatures are known for their use in anonymous credential systems to prove membership without revealing identity \cite{IF2023}. In a CBDC, BBS+ signatures are proposed for the creation of coins for the benefit that the group manager (the CB) can, when required, reveal information~\cite{z05}. 

\subsubsection{Blind Signatures}\label{sec:blind}
For a CBDC to achieve a digital equivalent of cash, the link between the issuing authority (the CB) and the user must be broken. This function can be accomplished with blind signatures~\cite{Chaum1983} that allow a signer to sign a message without viewing its content, ensuring the signer remains ignorant of the message while still vouching for its authenticity. This is useful for digital cash systems to issue currency while maintaining user privacy and preventing double spending. 

Blind signatures are applicable in the context of CBDCs \cite{z68,z74,z51,z50}, and seem promising. A bank can use blind signatures to sign electronic tokens representing currency units without seeing the actual token's details. When a user spends the digital cash, the bank cannot link the creation of the token to its subsequent spending, ensuring user privacy similar to physical cash. They can be used in certain protocols to ensure privacy or in off-chain solutions where transaction details need to be obscured from certain participants. The main drawback is that the user cannot audit the message they are signing. This drawback can be analogised by signing the outside of a sealed envelope attesting that it was in your possession, however, you did not open and read it.

The scheme by Kiayias et al. \cite{z51} uses blind signatures to protect user privacy. This method preserves anonymity and security. The maintainers know they have signed a valid transaction or account snapshot, but they don't know the specific details of what they signed. The user, on the other hand, ends up with a legitimately signed document without exposing sensitive transaction details to each individual maintainer. Chu et al. \cite{z74} highlight the use of blind signatures in offline transaction capability such as with smart cards.

\subsubsection{Verifiable Random Functions}
The last digital signature variant, a Verifiable Random Function (VRF), can serve as a proxy for a digital signature. Both VRFs and digital signatures involve the use of private keys to produce some output (a signature or a random value) and the use of corresponding public keys to verify that output. The VRF produces a random output from a given input in a deterministic manner, where the output can be verified with a proof by anyone holding the corresponding public key. A user with a private key can compute a random value from an input using the VRF, and a proof. Anyone with the public key can then verify, using the proof, that the random value was generated correctly without knowing the private key. A VRF can be seen as a special type of signature scheme where the \textit{signature} is a random value that is deterministically derived from an input and can be verified with a proof. 

Lee et al. \cite{z50} mention Groth signatures\footnote{Groth signatures \cite{Groth2010} used as part of ZK proving, provide a way to produce short signatures and are particularly efficient in terms of verification.} and VRFs as part of a method to overcome the trusted setup assumption present with SNARKs (\cref{sec:zk}) although this is in the context of permissioned blockchains and not CBDCs \cite{Androulaki2020}.

\subsection{Multi-Party Computation}
Multi-Party Computation (MPC) and commitment schemes such as Pedersen Commitments (PC; \cref{sec:pc}) both deal with hiding information. MPC, based on secure two-party computation \cite{Yao1982}, is designed to compute functions over secret data without revealing the data. Multiple parties can come together, each bringing their inputs, then MPC can compute a result, and the individual parties are not privy to each other's data. This ensures data privacy, even when collaborating on computations (Layer C in \cref{fig:layers}). One of the primary applications of MPC in cryptocurrencies is for secure wallet generation and key management. Traditional single-key wallets face risks if the key is lost or stolen. Using MPC, a private key can be split into multiple shares distributed among various parties. To sign a transaction, a threshold number of these shares are required. This approach adds a layer of security, making it harder for attackers to compromise a wallet since they would need to gain access to multiple key shares. A version of this has been implemented in a CBDC construction on Cosmos \cite{z06}. 

MPC is similar to Homomorphic Encryption (HE) (\cref{sec:he}) but under different contexts: MPC requires multiple parties to interact and be collaborative, whereas HE focuses on calculations of encrypted data. For transfers of digital currency, MPC can be applied to obfuscate details like the transaction amount while still allowing for network validation. Monero calls these confidential transactions. Transaction aggregation is a similar technique that can benefit MPC, given all parties agree to the aggregation. For example, Lee et al. \cite{z50} applies MPC to protect privacy during real-time gross settlement between banks.

\subsection{Pedersen Commitments}\label{sec:pc}
Pedersen Commitments (PCs) \cite{Pedersen1992} provide a secure mechanism for committing to a message by encapsulating it within a cryptographic envelope. This approach ensures the confidentiality of the original message, safeguarding it from disclosure until the committing party opts to unveil it. The robustness of this technique stems from its binding property, which guarantees that once a message has been committed, any attempt to alter it will be computationally detectable, thereby preserving the integrity of the commitment. Wang et al.~\cite{z30-53} apply a Pedersen commitment to allow for auditability of commercial banks by the central bank.

PCs are additively homomorphic meaning that the commitments can be added together without knowing the individual components which is applicable to preserving transaction information. They are also applied in ZKP scenarios to enable transactions that are both private and verifiable. The commitment allows a user to prove that they have knowledge of a value (like a secret key or transaction amount) without revealing the value itself. PCs are used to create transactions where the sender and receiver know the transaction details, but to the rest of the network, these details remain private (Layer C in \cref{fig:layers}) \cite{z10-49}. More broadly, outside of currency, they can be used to commit knowledge of a vote, or value of an asset, without revealing the vote recipient or value amount.

Androulaki et al.~\cite{Androulaki2020} propose an auditable anonymous token management system that uses PCs to conceal UTXOs (see \cref{sec:account}) in a permissioned blockchain. Project Khokha from the South African Reserve Bank \cite{SARB2018} applies Pedersen commitments and range proofs for transaction privacy in a permissioned instance of Quorum \cite{z15}. Both these projects highlight Layer E in \cref{fig:layers} to ensure compliance with regulation.

A different way of aggregation is found in the Mimblewimble protocol~\cite{Jedusor2016} which applies PCs to combine transactions into a single larger transaction so that an observer cannot determine the link between sender and recipient. Monero and Zcash also apply Pedersen commitments.

\subsection{Homomorphic Encryption}\label{sec:he}
Homomorphic Encryption (HE) \cite{Gentry2009} is a form of encryption that allows computations on ciphertexts, which, when decrypted, match the result of the operations as if they were performed on the plaintext. Essentially, it lets you work with encrypted data without decrypting it first. The primary allure of HE is secure data analysis. It allows for private computations in the cloud where the cloud server doesn't know the actual data it's computing on. Applications include encrypted search, privacy-preserving medical research, and secure voting systems. HE could enable a system where transaction details are encrypted, yet computations (such as verifying the transaction's authenticity or ensuring there are sufficient funds) can be carried out on these encrypted transactions. This means the transaction can be validated without the verifying entity ever seeing the transaction's specifics or the identities of the involved parties (Layer C in \cref{fig:layers}). Personal privacy can be preserved when CBs or other regulatory entities analyse transaction trends for monetary policy or anti-fraud measures without directly accessing the personal details of the involved parties. 

The downside is computational overhead making large volumes of data analysis impractical. Lee et al. \cite{z50} talk of HE for private calculations such that transaction amounts can be obscured while still adhering to transaction input and output balancing. Similar HE use is seen in the Wang et al.~\cite{Wang2020} framework for transaction privacy in Bitcoin.

To enhance user privacy while monitoring for illicit activities, the Australian Transaction Reports and Analysis Centre (AUSTRAC) collaborated with participants from the Fintel Alliance, a public-private initiative it founded, to create advanced data-matching and machine-learning tools aimed at detecting anomalous behaviour. Utilising HE, these tools can analyse and process machine-learning data while it remains in an encrypted state, thus preserving confidentiality \cite{z66}; Layers D and E in \cref{fig:layers}.

\subsection{Zero-Knowledge Proofs}\label{sec:zk}
The most frequently cited PET for enhancing CBDC privacy is Zero-Knowledge Proofs~\cite{z05,z11,z12,z15,z19,z34,z44,z50,z51,z56,z57,z69,z71,z10-49}. ZKPs enable one party (the prover) to demonstrate to another party (the verifier) that a statement is true, without disclosing any additional information about the statement itself~\cite{Goldwasser1989}. Within the context of CBDCs, ZKPs play a dual role: at Layer C (\cref{fig:layers}), they provide users with a mechanism to maintain anonymity during transactions, and at Layer D, they enable users to control access to their transaction data. ZKPs can ensure transaction anonymity by concealing sensitive details such as transaction amounts or user identities. Additionally, ZKPs facilitate selective disclosure, allowing specific information to be verified without revealing the underlying data—for example, proving that an account balance meets a certain threshold without disclosing the actual balance.

Zero-Knowledge Succinct Non-Interactive ARguments of Knowledge (SNARKs)~\cite{BenSasson2014} were introduced as a method to produce ZK proofs that are both succinct (compact in size) and non-interactive (requiring only a single communication round between the prover and verifier). In the realm of CBDCs, Gross et al.~\cite{Gross2021} propose leveraging SNARKs to enable private transactions within predefined limits, drawing inspiration from their implementation in privacy-focused cryptocurrencies like Zcash~\cite{z19,z10-49}.

While SNARKs achieve proof of computation, they do not inherently offer privacy--both the prover and verifier are aware of what is being proved. By incorporating a ZK component, SNARKs can ensure proof of computation without revealing the underlying data being verified. A classic example illustrating this concept involves solving a Sudoku puzzle: the prover demonstrates that they have solved the puzzle correctly, while the verifier can confirm the solution without seeing the specific puzzle answer~\cite{Berentsen2023}. This mechanism allows for verifiable computation while preserving privacy.

A notable limitation of SNARKs is their reliance on a trusted setup, which introduces potential vulnerabilities in generating the initial cryptographic parameters. Despite this drawback, zk-SNARKs have been successfully implemented in cryptocurrencies like Zcash, where they enable secure ledger maintenance by concealing transaction amounts and participant identities. Zero-Knowledge Scalable Transparent ARguments of Knowledge (STARKs)~\cite{BenSasson2018} were developed as an alternative to SNARKs, addressing the need for transparency by eliminating the requirement for a trusted setup. STARKs achieve cryptographic security through hash functions rather than number-theoretic assumptions like factorisation, making them resistant to quantum computing threats. However, STARK proofs are larger in size compared to SNARK proofs, which can impact efficiency. Ongoing research is focused on optimising STARK protocols to reduce proof sizes and improve computational performance.

Although STARKs have not been widely discussed in CBDC literature reviews, emerging technologies such as Starknet--a Layer-2 Ethereum rollup protocol--demonstrate potential infrastructure for supporting digital currency development. This represents an area of opportunity for CBDC research as such protocols continue to evolve and optimise.

\section{Privacy-adjacent Techniques}\label{sec:other}
A final category of techniques and considerations come from the blockchain era that are not traditional PETs, but closely related to design and implementation elements that can enhance privacy: the ledger model, coin mixing, secure hardware, and digital identity.

\subsection{Ledger Models}\label{sec:account}
An Unspent Transaction Output (UTXO) model treats a transaction as composed of inputs and outputs. For example, should the input be 10 coins to purchase services valued at 9 coins, two outputs are created: 9 coins to the merchant and 1 coin change to the payer. This model is advantageous for auditability, scalability, and security, but requires fresh addresses to improve privacy by removing the implicit user-address link. This contrasts with an account model that tracks a balance and a token model that allows for asset transfer in addition to native currency transactions. 

Islam and In~\cite{z09-67} design a UTXO-model CBDC in the context of a permissioned blockchain. To overcome the privacy need for dynamic address use they employ a centralised certificate authority (the CB) to manage IDs and the link to the wallet addresses. Repeated use of the same address can lead to de-anonymisation, so Pocher and Veneris~\cite{z10-49} use hierarchical deterministic wallet structures (a different address for each transaction) preventing the easy association of public addresses with individual users. Another proposal is a CBDC via a hybrid scheme to use an account model for basic transactions and a UTXO model for assets other than the native currency~\cite{z17}. This hybrid model is conceived to improve efficiency when considering a large number of transactions\footnote{According to data from the China National Network Clearing Corporation, the number of online transactions from January 24 to 30 during the Spring Festival holiday in 2020 was 4.919 billion \cite{z17}.} in a permissioned blockchain.
\subsection{Mixing}\label{sec:mixing}
Mixing is related to the idea of Privacy Pools (PP) which hide transactions in a crowd by obfuscating their links in a public ledger. Tornado Cash is a non-custodial privacy solution on Ethereum that breaks the on-chain link between source and destination addresses using zk-SNARKs. This is susceptible to tainting if a known malicious transaction is in the set; and by extension if the transaction originates from a tainted contract. Using ZK, Buterin et al.~\cite{Buterin2023} suggest an inverse protocol via PP, allowing users to disassociate from a transaction set that has been tainted by publishing a proof that their transaction did not originate from an identified source. 

Centralised mixing presents a trusted source that has access to the input transaction information, this is beneficial for auditing purposes and suggested by Pocher and Veneris~\cite{z10-49}, while decentralised mixing, such as Tornado Cash, does not allow for straight-forward auditing. Samourai, a Bitcoin wallet, has a feature called Whirlpool, that provides transaction mixing via CoinJoin to combine multiple transaction inputs into a single transaction. By doing this, it becomes significantly more difficult to link inputs to outputs.
\subsection{Secure Hardware}
Secure hardware refers to specially hardened computing infrastructure. Hardening can occur at the processor level in the form of a trusted execution environment (TEE) which is a separate enclave not accessible to other parts of the processor and computing stack, or as a separate module entirely combined of trusted software, firmware, and a chip \cite{Allen2020}. The TEE is more common, with smartphones capable of supporting them, for example, Samsung's KNOX. Standalone modules include hardware wallets for cryptocurrency storage, and smart cards that can use NFC and RFID technology, similar to debit cards \cite{Veneris2021}. Both TEEs and modules can be part of a CBDC design. 

A TEE can be leveraged by sending encrypted transaction data to a server to compute the expensive cryptographic operations, removing the computation burden on a mobile device. Depending on the cryptographic operations, user privacy can be enhanced~\cite{Allen2020}.

A secure cash card or NFC based system can allow for offline payments without contacting the bank~\cite{z50}. The bank originally issues the digital tokens which can be loaded into a secure card or mobile wallet. This yields cash-like privacy for the users and the benefit of transacting in rural areas, or as tourists outside local jurisdictions~\cite{Veneris2021}.

\rev{The digital euro bears this out at scale. In preparing for offline functionality, the ECB worked with device manufacturers and service providers to evaluate secure-environment form factors, identifying three candidates capable of meeting the tamper-resistance requirements of offline payments: embedded secure elements (eSEs), integrated secure elements (tamper-resistant enclaves within the device hardware, explicitly excluding TEEs), and embedded SIMs (eSIMs) \cite{ECB2025}. The explicit exclusion of TEEs is interesting: for offline money, where the device itself must prevent double spending, the bar is set above the isolation that a TEE provides.}

\subsection{Digital Identity}\label{sec:id}
Digital identity is closely tied to digital methods of financial participation. Cryptocurrencies like bitcoin and ether\footnote{A style note: coins are lowercase, and networks like Bitcoin and Zcash are capitalised.} are pseudonymous meaning that identity can be linked via transaction metadata analysis. De-anonymisation of financial activity leading to someone's past history is irreversible and can have downstream social effects. Although this review focuses on CBDCs, there is scope for brief discussion in the identity space.

De Portu~\cite{z20} suggests that in order to transact, a payee need only verify their personhood. Once they are identified, the necessary balance requirements can be verified, but that can be done after ID verification. The authors further suggest that NFTs can be the basis for a payee's digital identity. The proposed system is only pseudonymous in theory as publicly readable transactions are at risk of de-anonymization. Takaragi et al. \cite{z71} propose a system to apply a delegatable anonymous credential (DAC) via Pedersen commitments and ZKPs to national ID cards. Their system allows a user to prove they have valid credentials without revealing the private information behind them. Scollan and Darling~\cite{z69} suggest decentralised identity (DID) to manage a user's privacy disclosure where and when appropriate as a middle ground between total user anonymity and state surveillance.

\rev{\section{Comparing PETs for CBDC Deployment}\label{sec:pet-compare}
The layered stack in \cref{fig:layers} organises PETs by the privacy requirement they serve; deployment decisions additionally require comparing the technologies against operational criteria. \cref{tab:pet-compare} evaluates each PET on five dimensions: the privacy guarantee provided, compatibility with regulatory obligations (AML/CFT visibility and auditability), computational overhead, deployment maturity, and suitability for offline payments. The assessments synthesise the preceding sections with recent CB evaluations: the Bank of Canada's staff assessment of PETs for CBDC, which concludes that PETs carry meaningful performance overheads, added system complexity, and in several cases limited effectiveness owing to their early stage of development \cite{BoC2025}, and the Bank of England--MIT study finding that pseudonymisation, ZKPs, and MPC could feasibly minimise data sharing with both the central bank and intermediaries, within limits \cite{BoEMIT2024}.}

\begin{table*}[t]
\rev{
	\begin{threeparttable}
	\caption{\rev{Comparison of PETs for CBDC deployment across privacy guarantee, regulatory compatibility, computational overhead, deployment maturity, and offline-payment suitability. The technologies with the strongest anonymity guarantees---ring signatures, mixing, and unconstrained ZKPs---score worst on regulatory compatibility. Layer refers to \cref{fig:layers}. Assessments are qualitative syntheses of \cref{sec:pet,sec:other}, and CB evaluations \cite{BoC2025,BoEMIT2024,ECB2025}.}}
	\label{tab:pet-compare}
	\fontsize{8pt}{10pt}\selectfont
	\begin{tabularx}{\textwidth}{
			@{} 
			>{\raggedright\arraybackslash\hangindent=1em\hangafter=1}p{2.6cm} 
			l 
			>{\raggedright\arraybackslash\hangindent=1em\hangafter=1}X 
			>{\raggedright\arraybackslash\hangindent=1em\hangafter=1}X 
			>{\raggedright\arraybackslash\hangindent=1em\hangafter=1}p{2.2cm} 
			>{\raggedright\arraybackslash\hangindent=1em\hangafter=1}p{2.2cm} 
			l
			@{}
			}
		\toprule
		\textbf{PET} & \textbf{Layer} & \textbf{Privacy Guarantee} & \textbf{Regulatory Fit} & \textbf{Overhead} & \textbf{Maturity} & \textbf{Offline} \\
		\midrule
		Standard Encryption & A & Data confidentiality only; no transactional privacy & High (expected baseline) & Negligible & Production & Yes \\
		Digital Signatures & B & Authenticity and integrity; pseudonymity at best & High (supports identity binding) & Negligible & Production & Yes \\
		Blind Signatures & C & Issuer--spender unlinkability (payer anonymity) & Medium; compatible with value caps & Low & Piloted (Tourbillon) & Yes \\
		Ring Signatures & C & Sender ambiguity within a set & Low; impedes tracing & Moderate (grows with ring size) & Production$^*$ & Poor \\
		BBS$+$ Signatures & C, E & Anonymity with designated-authority de-anonymisation & High (accountable anonymity) & Low--moderate & Standardisation in progress & Moderate \\
		MPC & C, E & Joint computation without sharing inputs & High (audit without data access) & High; interactive & CB experiments & No \\
		Pedersen Commitments & C, E & Hides amounts; binding; auditable with range proofs & Medium--high & Low & Production$^*$; CB experiments & Limited \\
		Homomorphic Encryption & C, D, E & Computation on encrypted data & High (analysis without access) & Very high & Research stage & No \\
		zk-SNARK & C, D & Selective disclosure; hides amounts and identities & Contested; failed AML tests in Drex \cite{Gomes2025} & High proving cost; trusted setup & Production$^*$; failed CBDC pilots & Poor \\
		zk-STARK & C, D & Same as SNARK; no trusted setup; post-quantum & Same as SNARK; untested in CBDC pilots & Larger proofs & Production$^*$ & Poor \\
		Secure Hardware (TEE) & -- & Hardware-enforced confidentiality; trust shifts to vendor & High (controlled environment) & Low at runtime & Production (smart cards, enclaves) & Yes \\
		Mixing / privacy pools & C & Unlinkability within a crowd & Low unless centralised or provable \cite{Buterin2023} & Low--moderate & Production$^*$ & No \\
		\bottomrule
	\end{tabularx}
	\begin{tablenotes}
		\item[$^*$] In production only in cryptocurrencies. 
	\end{tablenotes}
	\end{threeparttable}
}
\end{table*}

\rev{Three observations follow from the table. First, the technologies with the strongest anonymity guarantees (ring signatures, mixing, unconstrained ZKPs) score worst on regulatory compatibility, while accountable-anonymity constructions (BBS+ signatures, auditable commitments, ZKPs constrained to thresholds) preserve a supervisory channel and are therefore the realistic candidates for deployment. Second, computational overhead remains a binding constraint: fully homomorphic encryption and on-device ZK proving are not yet practical at retail-payment scale \cite{BoC2025}. Third, offline payments, a stated requirement in several designs \cite{England2025offline,ECB2025}, are served almost exclusively by secure hardware, which provides the weakest cryptographic privacy guarantee of the set; the ECB's offline digital euro work, centred on secure-element form factors rather than cryptographic protocols, corroborates this pattern \cite{ECB2025}. A CBDC requiring both offline capability and strong anonymity must therefore compose multiple PETs rather than select one.}

\section{Case Studies in CBDC Privacy Technology}\label{sec:case-studies}
Select CBDC implementations are chosen to evaluate their assessment and use of PET, as detailed in \cref{tab:case-studies}. The selection includes all projects currently in the launched phase, as identified by the Human Rights Foundation during the course of this research. Additionally, pilot-phase projects are included to provide a representative sample of initiatives that have been made available to citizens at some stage. To ensure balance, high-profile proof-of-concept (PoC) and research projects are incorporated to emphasise the global scale of CBDC development. While the boundaries between these categories are not rigid, PoC projects are defined as those with design documents that could feasibly be implemented, whereas research-only projects remain at the desktop research or working group stage. Countries that have cancelled their CBDC initiatives, such as Denmark, Kenya, and Canada, are excluded from this analysis and discussed in \cref{sec:disc}.

\begin{table*}
	\begin{threeparttable}
	\caption{Case-studies in PET organised by project status and ordered by project age (most mature at the top). The trend is for CBDCs to explore privacy technology at the research stage which gets pared back or eliminated by the time the project launches. Layer maps to the cryptographic stack in \cref{fig:layers}, for example, \textbf{C} indicates PET to enhance user anonymity, with the distinction \textbf{C*} for assumed functionality because the details are not specified. DLT is included to highlight the large number of permissioned Hyperledger implementations.}
	\label{tab:case-studies}
	\fontsize{8pt}{10pt}\selectfont
		\begin{tabularx}{\textwidth}{
				@{}
				>{\raggedright\arraybackslash\hangindent=1em\hangafter=1}p{4cm}
				>{\raggedright\arraybackslash\hangindent=1em\hangafter=1}p{4.7cm}
				>{\raggedright\arraybackslash\hangindent=1em\hangafter=1}p{4.7cm}
				l
				X
				@{}
			}
			\toprule
			\textbf{Central Bank (CBDC)} & \textbf{PET Approach} 		& \textbf{Anonymity Features} 			& \textbf{Layer} & \textbf{DLT} \\
			\midrule
			\multicolumn{4}{l}{\textbf{\textit{Status: Launched}}} \\
			\midrule
			China (eCNY) 			& Internal firewall 			& Managed anonymity for small transactions 	& C* & No \\
			Bahamas (Sand Dollar) 	& Encryption, two-tier model 			& Reduced KYC for lower tier 		& B & Private \\
			Nigeria (eNaira) 		& Encryption \& multi-factor authentication, two-tier model & Reduced KYC for lower tier & B & HL Fabric$^1$ \\
			Jamaica (JAM-DEX) 		& Encryption, digital signatures, \& multi-factor authentication	& No anonymity	& B	& No \\
			Iran (Digital Rial) 	& Not specified 						& Not specified 					& B & HL Fabric \\
			Russia (Digital Ruble) 	& FSB-certified cryptography 			& No anonymity						& B & Hybrid \\
			Kazakhstan (Digital Tenge) & Two-tier model 					& Customisable anonymity			& C* & Corda R3$^2$ \\
			\midrule
			\multicolumn{4}{l}{\textbf{\textit{Status: Pilot}}} \\
			\midrule
			ECCB$^5$ (DCash) 	& Two-tier model 						& No anonymity							& B & HL Fabric \\
			Ghana (eCedi) 			& Two-tier model 						& No anonymity						& B & No \\
			India (Digital Rupee) 	& Not specified							& No anonymity						& B* & Not specified \\
			Brazil (Drex) 			& ZKP tests failed AML compliance 		& No anonymity						& B & HL Besu$^3$ \\
			Solomon Islands (Bokolo) & Not specified 						& Not specified 					& B & HL Iroha$^4$ \\
			Thailand (Digital Baht)	& Identity/payment data separation		& Not specified 					& B* & Undecided \\
			\midrule
			\multicolumn{4}{l}{\textbf{\textit{Status: Proof of Concept}}} \\
			\midrule
			ECB (Digital Euro) & Identity/payment data separation, pseudonymised settlement & \rev{Offline cash-like privacy via on-device secure element} & C & \rev{No$^6$} \\
			France$^7$ 		& ZKP, Two-tier model					& Possible 							& C & HL Fabric \\
			Boston Fed/MIT (Project Hamilton) & UTXO model, 2PC consensus 	& Possible 							& C & No \\
			Japan (Digital Yen) 	& Identity/payment data separation, UTXO model, TEE, MPC, HE 	& Possible 	& E & Undecided \\
			\midrule
			\multicolumn{4}{l}{\textbf{\textit{Status: Research}}} \\
			\midrule
			Norway 					& Not specified 						& Not specified 					& -- & HL Besu \\
			UK (Digital Pound) & ZKP, MPC, two-tier model 				& No anonymity							& E & Not specified \\
			Switzerland$^8$	& Blind Signatures, mixing (Tourbillon) & Payer anonymity (Tourbillon), No anonymity (Helvetia) & C & HL Fabric \\
			\bottomrule
		\end{tabularx}
	\begin{tablenotes}
				\item[$^1$] HL is Hyperledger Fabric, a modular enterprise blockchain framework. Fabric, Besu, and Iroha are all DLTs by the Linux Foundation Decentralized Trust;
				\item[$^2$] Corda R3 is a permissioned DLT;
				\item[$^3$] Besu is an enterprise Ethereum client;
				\item[$^4$] Iroha is a general purpose enterprise DLT;
				\item[$^5$] The Eastern Caribbean Central Bank includes: Anguilla, Antigua and Barbuda, Commonwealth of Dominica, Grenada, Montserrat, Saint Kitts and Nevis, Saint Lucia, and Saint Vincent and the Grenadines;
				\item [$^6$] \rev{Earlier ECB experimentation used Corda R3; the 2025 design is a centralised ledger informed by DLT design principles.}
				\item[$^7$] France's experiments are in wholesale CBDC.
				\item[$^8$] Switzerland is researching two projects: Tourbillon (retail) and Helvetia (wholesale).	
	\end{tablenotes}
	\end{threeparttable}
\end{table*}

A key trend observed across the case studies is the extensive research into privacy technologies, which often fails to translate into practical implementation. Among launched CBDCs, China's eCNY represents the largest-scale initiative. It advertises a concept called \textit{managed anonymity}, purportedly designed to mimic cash-like anonymity for small transactions. However, the specifics of this system remain undisclosed, making it difficult to verify who has access to transaction data or how the managed anonymity mechanism operates. Additionally they admit that ``full anonymity for all transactions cannot be considered''~\cite{IMF2022}. Other launched projects, such as Nigeria's eNaira and the Bahamas' Sand Dollar, offer tiered anonymity models where KYC requirements are reduced for smaller transactions. Despite these measures, participation in these systems still requires some form of identification, such as a phone number, meaning that true cash-like anonymity is not achieved \cite{IMF2023}. Kazakhstan's Digital Tenge claims customisable anonymity at the request of the user, however there are no details about how it is achieved \cite{Kazakhstan2023}.

The lack of robust anonymity features is further emphasised by other launched projects, such as Jamaica's JAM-DEX and Russia's Digital Ruble, which offer no anonymity at all; similar to all six pilot-status CBDCs. While Russia advertises privacy via a ``FSB [Federal Security Service]-certified cryptographic information protection facility'' for its Digital Ruble, the specific cryptographic protocols remain unspecified, and are likely standard encryption \cite{Stepanchenko2022}. Similarly, China's privacy approach relies on an \textit{internal firewall}, essentially a cybersecurity system designed to protect data rather than ensure user privacy \cite{JiangLucero2023}. This conflation between data security and user privacy is a recurring theme across many implementations; while encryption is widely employed to secure data, it does not inherently provide transactional privacy. Six studies mention some form of a \textit{two-tiered} model in their privacy approach and a further three note the separation of identify information from payment transaction data. This model is employed to insulate the CB from citizen identity data, placing the burden on the middle-man, usually the retail bank. Unfortunately this does not assist users' privacy of financial transaction data, it just shifts data security downstream, closer to the user.

Pilot-phase initiatives reveal experimental approaches to privacy that have yet to be fully operationalised. For instance, Brazil's Drex explored ZKP during its testing phase but ultimately abandoned this technology due to challenges in meeting AML compliance requirements \cite{Brasil2023}. \rev{The first pilot phase tested privacy solutions intensively (including Anonymous Zether, EY's Starlight, and Parfin's Rayls)\footnote{Zether is an account-based confidential payment scheme for EVM-compatible ledgers using ZKPs; its anonymous variant conceals sender and receiver within a set of participants. Starlight is EY's open-source compiler that transforms a Solidity smart contract into a privacy-preserving equivalent enforced by ZKPs. Rayls (Parfin) takes an architectural approach: each participant operates a private EVM-based ledger, interoperable through a common communication layer, so only the ledger owner has visibility into its operations.}, and the central bank's account of the results is unusually candid: Zether anonymised transactions but ``did not guarantee full disclosure to regulators and other authorities,'' while Rayls preserved anonymity at a prohibitive scalability cost, processing fewer than 10 transactions per second against 150 for Drex without a privacy layer; in all cases response time, user experience, and programmability degraded \cite{Gomes2025}. The Deputy Governor characterised the outcome as a \textit{privacy trilemma} between privacy, scalability, and programmability, and the ongoing second phase has shifted focus from privacy to programmability and business models \cite{Gomes2025}. This episode is examined as a failure case in \cref{sec:failure}.} 

Similarly, advanced PETs such as UTXO models appear in proof-of-concept initiatives like Project Hamilton\footnote{A collaboration between the Massachusetts Institute of Technology and the Boston Federal Reserve} \cite{Hamilton2022} and the Digital Yen\rev{, and the ECB explored a UTXO model in its early experimentation}. The UTXO model is particularly promising for obscuring transaction identities; however, it has not yet been implemented in any launched or pilot-phase CBDCs. \rev{The ECB's current approach instead separates digital identity data from digital euro payment data, essentially a two-tier model. Complementing this, an IMF working paper on privacy protection for the digital euro discusses ZKPs and blind signatures to shield user data from improper use by the central bank \cite{IMF2024}, together with \textit{anonymity vouchers} for limited transactions, a concept originating in the ECB's 2019 proof of concept whose details were never specified \cite{ECB2019}.}

\rev{The digital euro has become the most advanced large-economy retail project: the Eurosystem\footnote{Includes the ECB and the euro-area national central banks.} concluded its two-year preparation phase in October 2025, with pilot exercises planned for 2027 and a possible first issuance in 2029 contingent on the EU legislative process \cite{ECB2025,ECB2025pr}. The closing report maintains the data-separation approach by settlement under pseudonymous identifiers such that the Eurosystem cannot connect a transaction to an individual. Cash-like privacy is also proposed for offline payments, although the cryptographic specifics remain unpublished \cite{ECB2025}. Concepts from the earlier experimentation, including anonymity vouchers, the UTXO model, and a Corda-based ledger, do not appear in the 2025 design, which instead adopts a centralised ledger informed by DLT design principles \cite{ECB2025}.}

Research stage projects demonstrate even more ambitious cryptographic experiments. For example, Switzerland's Tourbillon retail CBDC explores blind signatures and mixing techniques to enhance user privacy \cite{Tourbillon2023}. The UK's digital pound mentions multi-party computation to protect the sharing of sensitive data \cite{England2023} and the Bank of Japan talks of using MPC and homomorphic encryption to limit the spread of personal information \cite{Japan2023}. \rev{The Bank of England has since deepened this line of work: a joint study with the MIT Digital Currency Initiative concluded that pseudonymisation, ZKPs, and secure multi-party computation could feasibly be applied to a digital pound to minimise data sharing with both the central bank and payment intermediaries \cite{BoEMIT2024}, and a series of 2025 design notes, including one on deferred offline payments \cite{England2025offline}, feed into a design blueprint expected in 2026 \cite{England2025}. Notably, the Bank states that a digital pound would be private but not anonymous, with identity verification retained for financial-crime prevention \cite{England2025}.} Despite these advancements in CBDC design, it remains uncertain whether such PETs will be integrated into pilot or launched products.

Transparency in CBDC implementations remains limited. Among existing projects, only Project Hamilton\footnote{\url{https://github.com/mit-dci/opencbdc-tx}} and Brazil's Drex\footnote{\url{https://github.com/bacen/pilotord-kit-onboarding}} offer publicly accessible source code. In contrast, other initiatives predominantly utilise permissioned DLTs, often based on customised versions of Hyperledger or centralised proprietary systems. Although Hyperledger's base framework is open source, its central bank-specific customisations deviate from the collaborative principles typically associated with open-source software development. While it is not inherently expected that central banks adopt open-source solutions, the absence of transparency regarding the protocols and algorithms employed to achieve privacy objectives may undermine trust in these systems.

The case studies reveal notable discrepancies between the theoretical aspirations of PET and their practical implementation within CBDCs. Although ongoing research continues to investigate cryptographic solutions aimed at improving user privacy, operational deployments are frequently constrained by regulatory requirements.

\rev{\subsection{Failure Analysis: Why PETs Do Not Survive to Launch}\label{sec:failure}
The research-to-launch disconnect documented above is not a single phenomenon; the case studies point to four distinguishable root causes.

\textbf{Regulatory non-negotiables.} The clearest evidence comes from Brazil. In Drex's first pilot phase, Anonymous Zether performed its anonymisation function---and in doing so could not guarantee the disclosure to regulators and authorities that the central bank regards as a legal obligation for AML monitoring, fraud prevention, and monetary control \cite{Gomes2025}. The failure mode here is inverse to our intuition regarding technology: it over-performed against a requirement that was never negotiable. Privacy designs that lack such a built-in supervisory channel (\cref{tab:pet-compare}) are structurally unable to clear this hurdle, whereas accountable-anonymity constructions such as BBS+ signatures or threshold-limited ZKPs were not what the off-the-shelf solutions provided.

\textbf{Computational overhead and immaturity.} The Bank of Canada's assessment found that the PETs offering the strongest guarantees impose performance overheads and system complexity that are unproven at retail-payment scale, with several technologies judged to be at too early a stage of development for deployment \cite{BoC2025}. Drex again quantifies the constraint: Brazil's conventional RTGS processes 300 transactions per second and the Drex platform 150, but with the Rayls privacy layer enabled, throughput fell below 10 transactions per second \cite{Gomes2025}. On-device ZK proving and fully homomorphic computation remain similarly distant from the conventional operations they would replace.

\textbf{Accountability and liability allocation.} Two-tier architectures dominate deployed systems (\cref{tab:case-studies}) in part because they assign data-protection liability to commercial intermediaries, an allocation regulators and central banks already know how to supervise. Cryptographic privacy disturbs this allocation: if no party can read transaction data, no party can be held operationally accountable for monitoring it, a problem the Bank of England resolves by declaring the digital pound private but not anonymous \cite{England2025}.

\textbf{Institutional incentives.} Finally, the power--interest asymmetry of \cref{fig:power-interest} operates inside the design process: every institutional participant bears direct costs if oversight fails (sanctions, FATF grey-listing, fraud losses), while the cost of weak privacy falls diffusely on citizens who are not at the table. Absent an internal champion with power, privacy features are the natural margin to cut when budgets, deadlines, or compliance reviews bind.

These four causes operate together rather than separately: a PET that fails on any one of them does not reach launch, however well it satisfies the rest. The practical implication for designers is that privacy outcomes improve not by selecting stronger cryptography in isolation, but by composing PETs that preserve a supervisory channel by construction---and by committing to publish the mechanism so the privacy claim is verifiable.}

\section{Discussion}\label{sec:disc}
CBDCs may have arrived in the blockchain era, but do not necessarily use a blockchain or require blockchain technology.\footnote{For an overview of blockchain technology and CBDCs, see \cite{z07}.} It is the digital cash nature of cryptocurrency that lends itself to the idea that CBDCs are a suitable blockchain application (\cref{fig:categories}). This is evident in the large number of Hyperledger implementations in \cref{tab:case-studies}. Research that mentions blockchain in the context of privacy often presents a tenuous link and ignores regulatory requirements, or suggests strong anonymity via ZKPs, or addresses the pseudonymous nature of public blockchains by introducing a private consortium, thus negating the benefits of a public ledger. 

The cryptographic stack in \cref{fig:layers} details the PETs required to meet the definition of privacy, and the case studies presented in \cref{tab:case-studies} show the trend for well-intentioned privacy-preserving research that falls short with the launch of the CBDC. Both the Bank of England and the Bank of Japan meet Layer E in \cref{fig:layers} by proposing tools such as MPC and HE to process user data without having explicit access to it. These are the only two case studies that meet the requirement. By the time projects get to the pilot stage there are limited options for citizens to transact anonymously. The two exceptions are from China and Kazakhstan, with both advertising anonymity for small transactions. These are rated C* because the details are not specified and cannot be verified. The remainder of the CBDCs launched provide standard encryption and digital signatures as part of data security practices, but do not stretch to use more advanced PET. 

In terms of adoption, there has been limited success in introducing a CBDC to the market. \cref{tab:case-studies} shows seven countries that have launched CBDCs. The three earliest to market are Nigeria (2021), The Bahamas (2020), and China (2019), and they have struggled to grow adoption. The Nigerian CBDC, the eNaira, has seen limited acceptance, as Nigerians have very low trust in their government monetary system and there is high competition among mobile money operators~\cite{Ree2023}. The eNaira is linked to individual bank accounts and thus does not help the unbanked population or provide cash-like privacy despite reduced KYC requirements for the lowest tier transactions.

The Bahamian Sand Dollar was the first CBDC to graduate past the testing phase and launch. Although in use since 2020, there are only \$2.5 million Bahamian dollars in circulation, the low adoption rates are attributed to the lack of merchant acceptance and insufficient public education~\cite{AFI2024}. The technology is based on a private DLT, and although citizens do not need to KYC for transactions under \$500 (up to \$1500/month), there is no PET to safeguard user activity or identity~\cite{AFI2024,LedgerInsights2024}.

The largest scale CBDC operation comes from People's Bank of China (PBOC) with their eCNY pilot beginning in 2019. \rev{By the end of November 2025, the eCNY had processed 3.48 billion cumulative transactions totalling 16.7 trillion RMB (approximately US\$2.37 trillion) \cite{Xinhua2025}.} Privacy-enhancing technology is present in the form of a managed anonymity model supposedly allowing pseudo anonymous transactions under 10,000 yuan; although the exact technical details such as the type of encryption are not public~\cite{Orcutt2023}. The eCNY also has offline peer-to-peer capability that can enhance privacy in a limited way, but these transactions are still reversible and wallets can be deactivated by the PBOC~\cite{JiangLucero2023}. Overall adoption is very low, with most people preferring the convenience of existing payment operators Alipay and WeChat Pay, in addition to the reticence resulting from privacy and security concerns~\cite{Li2024}. \rev{Domestic adoption notwithstanding, the PBOC has shifted emphasis toward internationalisation, opening an e-CNY International Operations Center in Shanghai in September 2025 to develop cross-border payment, blockchain service, and digital asset platforms \cite{PBoC2025}; the transparency questions raised above remain unaddressed in this expansion. The most telling shift takes effect in January 2026: the eCNY becomes an interest-bearing, deposit-like instrument, abandoning its original positioning as a pure cash substitute \cite{Xinhua2025,Caixin2025} and, with it, the cash-like framing on which its managed anonymity was premised.}

Other countries involved in CBDC research and development have changed tack or outright cancelled their plans for broad introduction. Japan, Denmark, Kenya, \rev{Canada, and the United States} provide illustrative examples. In July 2022, the Bank of Japan (BOJ) released a report indicating that it had no immediate plans to issue a CBDC, primarily due to the ``strong preference for cash'' among the Japanese population. However, the BOJ established a CBDC Forum in July 2023, aimed at integrating expertise from various industry sectors to assist in the continued development of their CBDC pilot. The BOJ's April 2024 report concluded that ultimately the implementation of a CBDC will be ``decided by discussions among the public''~\cite{Japan2024}. 

Denmark's CBDC, the e-krone, has not proved itself worthy as it would not improve the already functioning financial infrastructure and would be in competition with their commercial banks. The National Bank of Denmark states ``it is not clear how a retail CBDC in Danish kroner can contribute to better and more secure access to payments and financial services''~\cite{Denmark2022}.

In Kenya, where the mobile payment service provider M-Pesa dominates the market, the Central Bank of Kenya has stopped research as of July 2023 concluding that the global ``allure of CBDCs is fading'' and ``implementation of a CBDC in Kenya may not be a compelling priority in the short to medium term,'' asserting that they will continue to monitor the landscape~\cite{Kenya2023}. 
%

The Bank of Canada has decided to shelve its plans for issuing a digital Canadian dollar after more than four years of research and exploration. This decision was announced in September 2024, marking a significant change in the central bank's approach to digital currencies. In addition to the negative public opinion mentioned in the Introduction (\cref{sec:intro}), the Bank of Canada Governor Tiff Macklem stated that there is currently no ``compelling case'' to proceed with a CBDC~\cite{Canada2024}. \rev{However, the Bank's research function has continued despite the shelving: its 2025  assessment of PETs for CBDC \cite{BoC2025} is among the most rigorous central bank treatments of the technologies surveyed in \cref{sec:pet}.}

\rev{The United States presents a categorical reversal. Executive Order 14178 of January 2025 prohibits federal agencies from establishing, issuing, or promoting a CBDC, citing risks to financial-system stability, individual privacy, and national sovereignty \cite{EO14178}. The House of Representatives passed the Anti-CBDC Surveillance State Act in July 2025, which would prohibit the Federal Reserve from issuing a retail CBDC or using one to implement monetary policy \cite{HR1919}. In parallel, the GENIUS Act of July 2025 established a federal framework for private payment stablecoins, signalling a policy preference for privately issued digital dollars over a central bank instrument. 
	
	The US case demonstrates that privacy is not only a design parameter within CBDC projects but can be the decisive political variable determining whether a project exists at all, reinforcing this paper's argument that verifiable, by-design privacy is a precondition for public legitimacy.}

\rev{Taken together, the trajectories tell one story from three directions. The CBDCs that launched without privacy have struggled to find users (China, Nigeria, the Bahamas); the projects that asked their citizens never launched (Canada, Japan); and the project that built its privacy architecture first (the digital euro) is the largest one still advancing. However a CBDC fares, the binding constraint is rarely cryptographic: the trajectories show that a CBDC must pass two tests that technology alone cannot clear, a value test against incumbent payment systems, which stopped Kenya and Denmark and stunts adoption in China and Nigeria, and a trust test with its own citizens, which stopped Canada and the United States and which the digital euro's privacy architecture is built to pass. Privacy cannot decide the first, but it anchors the second.}

\rev{\subsection{Limitations}\label{sec:limitations}
Three limitations bound the claims of this study. First, the case study evidence rests on publicly available documents, and many central banks do not disclose implementation details. Where a project advertises a privacy property without specifying its mechanism (for example, the eCNY's \textit{managed anonymity}), the property cannot be independently verified; such entries are explicitly marked as assumed (the asterisked layers in \cref{tab:case-studies}) rather than confirmed. The analysis is therefore an assessment of what central banks disclose, which is itself a finding (\cref{sec:intro}), but it cannot rule out undisclosed protections or undisclosed surveillance. Second, although every record was coded by one author and cross-reviewed by the other, no formal inter-coder reliability statistic was computed; the shared coding template, full cross-review, and resolution-by-discussion procedure substantially mitigate, but do not eliminate, subjectivity. Third, the layered privacy framework (\cref{fig:layers}) is evaluated diagnostically against 20 case studies, consistent with the demonstration step of DSRM, but it has not been validated in a working implementation. Constructing a proof-of-concept CBDC that instantiates the framework is left for future work (\cref{sec:future-work}).}

\rev{\subsection{Future Work}\label{sec:future-work}
Two directions for future work follow directly from the limitations acknowledged in \cref{sec:limitations}. First, the layered framework of \cref{fig:layers} should be validated constructively: a proof-of-concept implementation (for instance, an accountable-anonymity design composing blind signatures for issuance, BBS+ credentials for tiered identity, and threshold-limited ZKPs for transaction privacy) would allow the framework's performance, scalability, and regulatory compatibility to be measured rather than asserted, directly addressing the failure causes identified in \cref{sec:failure}. Second, research should move beyond reviewing what central banks choose to publish, toward auditable-by-default architectures in which privacy properties are publicly verifiable from open specifications and reproducible builds; such designs would convert the transparency deficit described in this paper from a research obstacle into a design principle.}

\subsection*{\rev{Concluding Remarks}}\label{sec:summary}
Cryptographic methods are recognised as integral to privacy (\cref{fig:categories}), with myriad options to enhance the user's right to maintain anonymity (Layer C in \cref{fig:layers}), however, they prove difficult to find in practice. The difficulty comes from allowing for a balance between privacy and auditability. Unlike many options to help with anonymity in Layer C, there are few options to help auditors comply with regulation in Layer D, and we observe no case studies that meet the criteria (although experiments by the BoJ and the Bank of England do suggest a path up to Layer E).

The majority of these case studies present at Layer B by employing standard encryption and digital signatures. They have little to no discernible privacy-enhancing technology; rather, they present as bureaucratic solutions without consideration of their constituents' needs. Perhaps they suffer from being too early to market without enough time for public opinion to sway in favour. In either case, there are few well-architected solutions for developers to learn from and build on. 

For fully articulated CBDC constructions, some are worth repeating here. Platypus \cite{z57} achieves privacy preservation within a centralised framework, while PEReDi \cite{z51} offers a distributed, privacy-preserving design that is resilient to single points of failure. Project Hamilton, represents a centralised proof-of-concept based on a UTXO transaction model. This model enhances efficiency and privacy by eliminating the need to store transaction data, though the authors acknowledge the inherent challenges in balancing citizen privacy with legal access \cite{Hamilton2022}. Similarly, Islam and In~\cite{z09-67} propose a consortium blockchain using a UTXO model to strike a balance between privacy, transparency, and auditability. In another proposal, Liu et al. \cite{z05} introduce an anonymous yet traceable CBDC; however, anonymity is limited to interactions between the central bank and commercial banks. Cos-CBDC \cite{z06}, built on the Cosmos-SDK framework, demonstrates a single blockchain prototype with claimed privacy-preserving properties. Although these academic projects illustrate promising applications of PET, they remain theoretical constructs with no practical implementation to date.

The importance of citizen rights and their perspectives on financial privacy highlights the need for further exploration of these potential solutions and new ones that can bridge the gaps. Achieving a balance between the interests of central banks and citizen stakeholders is crucial in the present age of digital finance including Central Bank Digital Currencies.

%

\section*{Author Contributions}
\textbf{Jeff Nijsse}: conceptualisation (equal); methodology (equal); investigation (equal); visualisation (equal); writing---original draft (equal); writing---review and editing (lead); funding acquisition (lead). \textbf{Andrea Pinto}: conceptualisation (equal); methodology (equal); investigation (equal); visualisation (equal); writing---original draft (equal).

\section*{Acknowledgements}
The authors would like to acknowledge Remo Nyffenegger for comments on an early manuscript and the anonymous reviewers for their helpful suggestions.

\section*{Funding Information}
This research was partially funded by the School of Engineering, Computer, and Mathematical Sciences at Auckland University of Technology.

\section*{Conflict of Interest Statement}
The authors declare no conflict of interest.

\rev{\section*{Data Availability Statement}
Data sharing is not applicable to this article as no new data were created or analysed in this study.}

\bibliography{cbdc-refs}

\end{document}